\documentclass[%
reprint,
superscriptaddress,
 amsmath,amssymb,
 aps,
]{revtex4-2}

\usepackage{mhchem}
\usepackage{graphicx}
\usepackage{dcolumn}
\usepackage{bm}
\usepackage{float}
\usepackage{placeins}
\usepackage{siunitx}
\usepackage{xcolor}
\usepackage{chemformula}
\setchemformula{kroeger-vink=true}
\DeclareMathOperator{\arcsec}{arcsec}
\usepackage{color,soul}
\usepackage[title]{appendix}

\begin{document}

\preprint{APS/123-QED}

\title{Engineering diamond interfaces free of dark spins}

\author{Xiaofei Yu}
\altaffiliation{These authors contributed equally to this work}
\affiliation{Department of Physics, The University of Chicago, Chicago, IL 60637}
\author{Evan J. Villafranca}
\altaffiliation{These authors contributed equally to this work}
\affiliation{Pritzker School of Molecular Engineering, The University of Chicago, Chicago, IL 60637}
\author{Stella Wang}
\affiliation{Department of Physics, The University of Chicago, Chicago, IL 60637}
\author{Jessica C. Jones}
\affiliation{Materials Science Division, Argonne National Laboratory, Lemont, IL, 60439, USA}
\author{Mouzhe Xie}
\affiliation{Pritzker School of Molecular Engineering, The University of Chicago, Chicago, IL 60637}
\affiliation{School of Molecular Sciences, Arizona State University, Tempe, AZ 85281, USA}
\author{Jonah Nagura}
\affiliation{Pritzker School of Molecular Engineering, The University of Chicago, Chicago, IL 60637}
\author{Ignacio Chi-Dur\'an}
\affiliation{Department of Chemistry, The University of Chicago,
Chicago, IL 60637}
\author{Nazar Delegan}
\affiliation{Materials Science Division, Argonne National Laboratory, Lemont, IL, 60439, USA}
\affiliation{Center for Molecular Engineering, Argonne National Laboratory, Lemont, IL, 60439, USA}
\author{Alex B. F. Martinson}
\affiliation{Materials Science Division, Argonne National Laboratory, Lemont, IL, 60439, USA}
\author{Michael E. Flatt\'e}
\affiliation{Department of Physics and Astronomy, University of Iowa, Iowa City, Iowa 52242, USA}
\affiliation{Department of Applied Physics, Eindhoven University of Technology, Eindhoven, 5600 MB, The Netherlands}
\author{Denis R. Candido}
\affiliation{Department of Physics and Astronomy, University of Iowa, Iowa City, Iowa 52242, USA}
\author{Giulia Galli}
\affiliation{Pritzker School of Molecular Engineering, The University of Chicago, Chicago, IL 60637}
\affiliation{Materials Science Division, Argonne National Laboratory, Lemont, IL, 60439, USA}
\author{Peter C. Maurer}
\email{pmaurer@uchicago.edu}
\affiliation{Pritzker School of Molecular Engineering, The University of Chicago, Chicago, IL 60637}
\affiliation{Materials Science Division, Argonne National Laboratory, Lemont, IL, 60439, USA}
\affiliation{CZ Biohub Chicago, LLC, Chicago, IL 60642, USA}




\begin{abstract}
Nitrogen-vacancy (NV) centers in diamond are extensively utilized as quantum sensors for imaging fields at the nanoscale. The ultra-high sensitivity of NV magnetometers has enabled the detection and spectroscopy of individual electron spins, with potentially far-reaching applications in condensed matter physics, spintronics, and molecular biology. However, the surfaces of these diamond sensors naturally contain electron spins, which create a background signal that can be hard to differentiate from the signal of the target spins. In this study, we develop a surface modification approach that eliminates the unwanted signal of these so-called dark electron spins. Our surface passivation technique, based on coating diamond surfaces with a thin titanium oxide (\ch{TiO_2}) layer, reduces the dark spin density. The observed reduction in dark spin density aligns with our findings on the electronic structure of the diamond-\ch{TiO_2} interface. The reduction, from a typical value of $2,000$~$\mu$m$^{-2}$ to a value below that set by the detection limit of our NV sensors ($200$~$\mu$m$^{-2}$), results in a two-fold increase in Hahn-echo coherence time of near surface NV centers. Furthermore, we derive a comprehensive spin model that connects dark spin relaxation with NV coherence, providing additional insights into the mechanisms behind the observed spin dynamics. Our findings are directly transferable to other quantum platforms, including nanoscale solid state qubits and superconducting qubits. 

\end{abstract}

\maketitle


\section{Introduction}

Electron paramagnetic resonance (EPR) spectroscopy has had far-reaching applications in chemical analysis \cite{JOSEPH1993926, B918782K, RESZKA200453}, biological \cite{WOLDMAN1994195, doi:10.1152/japplphysiol.01024.2002, doi:10.1021/acs.jpcb.2c05235} and medical research \cite{doi:10.1073/pnas.91.8.3388, clinicalEPR}, material science \cite{REX19872134, HARVEY198911, Felton_2009, doi:10.1021/cm400728j}, and condensed matter physics \cite{10.1063/1.1680446, PhysRevB.101.245204}. 
Although a versatile tool, EPR spectroscopy inherently suffers from low sensitivity, resulting in a typical state-of-the-art detection limit on the order of $10^{9}$ spins \cite{BrukerEPR}. 
Owing to their nanoscale size, NV centers in diamond can significantly improve the limit of detection, and in principle achieve single-molecule sensitivity. 
For example, NV centers have enabled spectroscopy on individual proteins \cite{doi:10.1126/science.aaa2253} and DNA \cite{NatMethods.15.697} molecules labeled with 2,2,6,6-tetramethylpiperidine-1-oxyl (TEMPO). 

However, applications beyond these proof-of-principle experiments have remained challenging, due to the presence of paramagnetic defects associated with the diamond surface. 
The signal from these so-called ``dark" spins can, in fact, be significantly stronger than that of the intended target, resulting in challenges in signal acquisition and interpretation. 
NV spectroscopy combined with a scanning magnetic tip has allowed the mapping of these dark spins, revealing that they are consistent with spins residing within a 1.5 nm thick layer on the diamond surface \cite{NatureNanotech.9.279}. 
A more recent investigation based on fitting the time evolution of measured Hahn-echo experiments suggest that these surface spins are mobile and have typical densities ranging from 1,000 to 50,000~$\mu$m$^{-2}$ \cite{PRXQuantum.3.040328}. 

Microwave control, such as dynamical decoupling \cite{PhysRevX.9.031052} and incoherent driving \cite{npjQuantumInf.8.47} of the surface spin, can efficiently decouple NV centers from dark spins. 
The dark spins have a g-factor of $2.0028(3)$ \cite{10.1063/1.5085351}, which is close to that of many target systems, including TEMPO-labeled molecules. 
As a result, spectral separation of a g$\approx$2 target spin from the dark spin bath would require several Tesla magnetic fields, making such an approach technically challenging. 
An alternative approach relies on reducing the dark spin density through material engineering.  
Advances in surface processing, cleaning, and annealing protocols have resulted in a modest reduction in dark spin density \cite{PhysRevX.9.031052}. 
Likewise, depleting surface charge traps using strong local electric fields provided by the charged tip of an atomic force microscope has enabled a 10-fold reduction in dark spins density to 300~$\mu$m$^{-2}$ \cite{NatPhys.18.1317}. 
However, this approach does not enable modification of large areas and is associated with significant experimental complexity, preventing a majority of EPR sensing applications. 
More recent experiments have shown that the deposition of a graphene layer on a diamond surface leads to a 2-fold reduction in dark spin density down to 1100~$\mu$m$^{-2}$ \cite{hao,lo2025enhancement}, but it remains unclear if this approach can be further improved.
Moreover, it remains unclear whether either of these techniques is compatible with chemical and biological functionalization for nuclear magnetic resonance (NMR), especially when compared with metal-oxide layer deposition \cite{liu2022surface}.
Earlier studies have also demonstrated that the density of paramagnetic surface spins can be artificially increased through specific surface treatments and molecular adsorption, providing valuable insight into the mechanisms governing surface spin formation and dynamics \cite{NewJPhys.13.055004, PhysRevB.86.195422}.

In this work, we demonstrate the creation of a nearly dark-spin-free surface by engineering tailored diamond-\ch{TiO_2} heterostructures, inspired by recent advances in diamond core-shell nanoparticles \cite{zvi}.
We find that the growth of \ch{TiO2} on diamond via atomic layer deposition (ALD) falls into two regimes: (1) inhibited growth for films formed with fewer than approximately 75 ALD cycles and (2) linear film growth for 75 or more ALD cycles.
We show that the two growth regimes demonstrate markedly different strengths of surface spin properties as well as NV spin coherence. 
To precisely evaluate the properties of dark spins following ALD, we developed a new pulse sequence that directly measures the dark spin relaxation rate while suppressing modulation from surrounding nuclear spins. Furthermore, we present a quantum model of dark spins that extends beyond previous studies by simultaneously incorporating their density, relaxation rate, proximity to NV centers, and bath dimensionality. Prior work relied on simplified models that neglected dark spin relaxation \cite{PRXQuantum.3.040328} or relied on classical filter function methods \cite{mamin2012detecting,davis2023probing}, limiting accuracy.
Collectively, our work yields a more accurate and robust characterization of dark spins situated at diamond surfaces. In addition, we performed an \textit{ab-initio} study of the diamond–TiO$_2$ interface, demonstrating the potential of using TiO$_2$ to passivate unsaturated dangling bonds on the diamond surface and analyzing the interfacial band alignment to evaluate whether the interface facilitates charge transfer from the diamond surface.

%

The paper is organized as follows: {\it Section A} describes the growth of the diamond-\ch{TiO_2} heterostructures and their material characterization. 
{\it Section B} focuses on quantum sensing-enabled spectroscopy to investigate surface dark spin resonance and NV coherence as a function of \ch{TiO_2} film thickness. 
{\it Section C}  experimentally examines the relaxation of the dark spins in these heterostructures. 
{\it Section D} presents a model that quantitatively captures NV dynamics as a function of dark spin density and relaxation, as well as distance. 
Finally, in {\it Section E} we perform \textit{ab-initio} calculations to understand the interfacial structure and band-alignment of the diamond-TiO$_2$ heterostructures. 

\section{Results}
\subsection{Diamond-\ch{TiO_2} heterostructures}

The samples utilized in this work are (100) single crystal electronic grade diamond substrates subjected to low-energy nitrogen implantation at a dose of $3 \times 10^{9}$~cm$^{-2}$ at 3~keV (Sample 0), $2 \times 10^{11}$~cm$^{-2}$ at 4~keV (Sample 1), $2 \times 10^{12}$~cm$^{-2}$ at 4~keV (Sample 2), and $1 \times 10^{13}$~cm$^{-2}$ at 5~keV (Sample 3). 
Implantation was done with \textsuperscript{14}N for Sample 3 while all others were implanted with \textsuperscript{15}N.
Sample 0 contains optically resolvable NV centers, while Sample 1 through 3 have NV densities that are not resolvable by diffraction-limited microscopy. 
Following ion implantation, the substrates were vacuum annealed at 800$^\circ$C according to established diamond processing protocols \cite{PhysRevX.9.031052}. 
We refer to Appendix \ref{Diamond Sample Processing} for more detailed information on diamond substrate processing.

\begin{figure}[ht]
\centering
\includegraphics[width=0.5\textwidth]{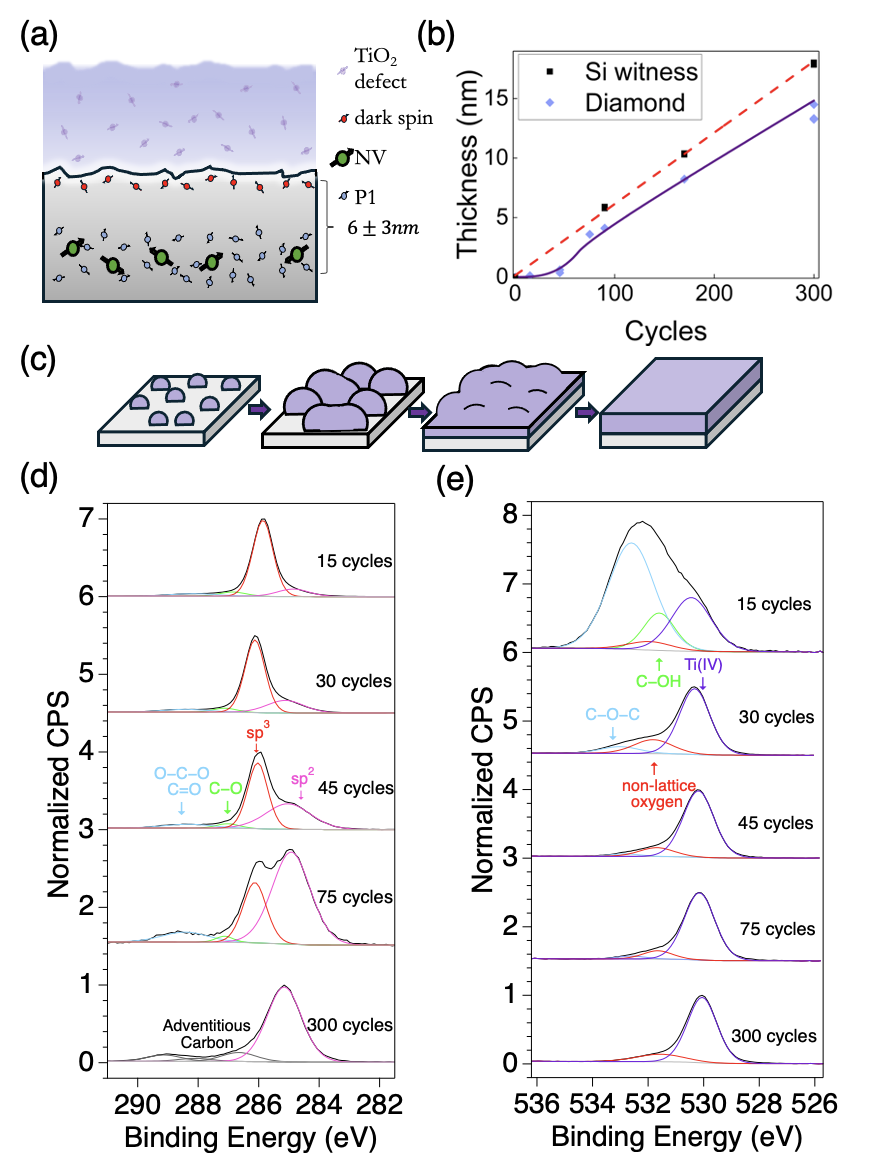}
\caption{Layout and characterization of ALD-coated \ch{TiO2} films on diamond.
(a) Schematic of \ch{TiO_2} heterostructures on diamond with implanted near-surface NV centers. (b) Ellipsometry measurements of film growth with varying numbers of ALD cycles to determine the resulting film thicknesses. Data is fit to the Nilsen equation describing the delayed nucleation on the diamond surface allowing for the density of nucleation sites to be estimated. (c) Illustration of the stages of the island growth model for site selective nucleation of ALD \ch{TiO_2} on diamond surfaces. (d) XPS carbon 1s spectra for varying number of ALD cycles. After 300 cycles only adventitious carbon incorporated into the \ch{TiO_2} film is measureable. (e) XPS oxygen 1s spectra for varying number of ALD cycles. Peaks are normalized to the Ti(IV) peak.}
\label{fig:MatCharFig}
\end{figure}

A schematic of the diamond-\ch{TiO2} heterostructures investigated in this study is given in Fig.~\ref{fig:MatCharFig}(a).  The \ch{TiO2} coatings were grown using ALD in a Savannah G2 chamber, employing alternating cycles of deionized \ch{H_2O} and tetrakis(dimethylamino)titanium (TDMAT) at a chamber temperature of 100$^{\circ}$C. 
The thickness of \ch{TiO_2} on the diamond surface was determined by \textit{ex situ} variable angle spectroscopic ellipsometry on a J.A. Woollam M2000 spectrometer. 
Using a Cauchy model for the diamond and \ch{TiO_2} overlayer, we extracted the \ch{TiO_2} thicknesses shown in Fig.~\ref{fig:MatCharFig}(b). 
Interestingly, for a small number of cycles (i.e., less than 50), the \ch{TiO_2} growth is delayed and we only see a linear growth for 75 or more cycles. 
This growth is in stark contrast to non-inhibited ALD growth on a native oxide silicon substrate as shown by the linear growth on silicon witness chips placed in the growth chamber along with our diamond substrates. 
The observed inhibited growth can be explained by an island growth model, where the precursor reacts with functional groups on the surface in the first few ALD cycles, forming sparse ALD nucleation sites.  
During the initial phase, the islands grow uniformly in all available directions with each subsequent cycle until the film coalesces to form a continuous conformal coating (Fig.~\ref{fig:MatCharFig}(c)). Following the merging of these islands, the film thickness increases linearly with the number of ALD cycles.
An analytical model developed by Nilsen et al. is used to fit the inhibited growth of our \ch{TiO_2} films ~\cite{Parsons, Nilsen} resulting in an estimated nucleation density of 30,000 $\pm$ 10,000 sites per $\mathrm{\mu m^2}$ (see Appendix \ref{INM}). 
This corresponds to an approximate island radius of 3 $\pm$ 1~nm required to achieve coalescence. 
Previous studies have suggested that water treatment of diamond surfaces under ALD conditions can alter nucleation rates and densities~\cite{Jones.Carbon, Nilsen, Choy}. 
We find that our \ch{TiO2} films demonstrate robust stability when immersed in solvent for multiple weeks and can be readily functionalized using the protocol from Ref. \cite{PNAS.119.8}, highlighting their applicability for quantum biosensing experiments (results summarized in Appendix \ref{TiO2Stability}). 

Next, we investigate the quality of the \ch{TiO_2} film  and corroborate the sparse nucleation sites for ALD on the diamond surface using X-ray photoelectron spectroscopy (XPS). 
As shown in Fig. \ref{fig:MatCharFig}(d), with increasing numbers of ALD cycles, carbon on the diamond surface singly bonded to oxygen decreases while the carbon doubly bonded to oxygen increases. 
A corresponding peak for ether-bonded carbon (C-O-C) occurring at 533 eV is also seen in the oxygen 1s spectra along with a peak for hydroxyl groups bonded to carbon on the diamond surface which becomes negligible after 15 cycles (Fig.~\ref{fig:MatCharFig}(e)).
This is consistent with island growth resulting from nucleation occurring primarily in the first few cycles on the sparse number of hydroxyl groups present in our nominally oxygen-terminated diamond substrates ~\cite{PhysRevX.9.031052, Jones.Carbon}. 
The total signal arising from the \ch{TiO_2} in the oxygen 1s spectra also follows the trend we expect for inhibited \ch{TiO2} growth. 
At 15 cycles, the oxygen peaks in \ch{TiO_2} account for $\sim 30 \%$ of the total signal.   
This percentage steadily grows with an increasing number of cycles as the islands coalesce and the photoelectrons no longer escape from the diamond surface under the \ch{TiO_2} film. 
We also observe an increase in \ch{sp^2} carbons with increasing cycles of \ch{TiO_2}, which we attribute to carbonaceous groups incorporated in the \ch{TiO_2} film from the incomplete reaction of the precursor during deposition, a process previously described in Ref. \cite{Dufond2020}. As the film fully coalesces, these adventitious carbon groups completely overwhelm the signal coming from the diamond surface. 
Indeed, as shown by the 300 cycle data in both the carbon 1s and oxygen 1s spectra, only photoelectrons coming from the \ch{TiO_2} film are measurable when the film is roughly 14 nm thick (see Supplementary Appendix \ref{XPS} and Supplementary Fig. \ref{fig:XPSsupp}). 
Additionally, our XPS results reveal that the resulting \ch{TiO_2} films are almost fully stoichiometric.
We are able to place an upper limit on the Ti(III) inclusion in our film at 7 ppk. and can conclude that this defect concentration is independent of the number of ALD cycles (see the fitting parameters for the titanium 2p spectra in Appendix \ref{XPS}).
%

\subsection{Spectroscopy on surface dark spins} 
\begin{figure}[ht]
\centering
\includegraphics[width=0.48\textwidth]{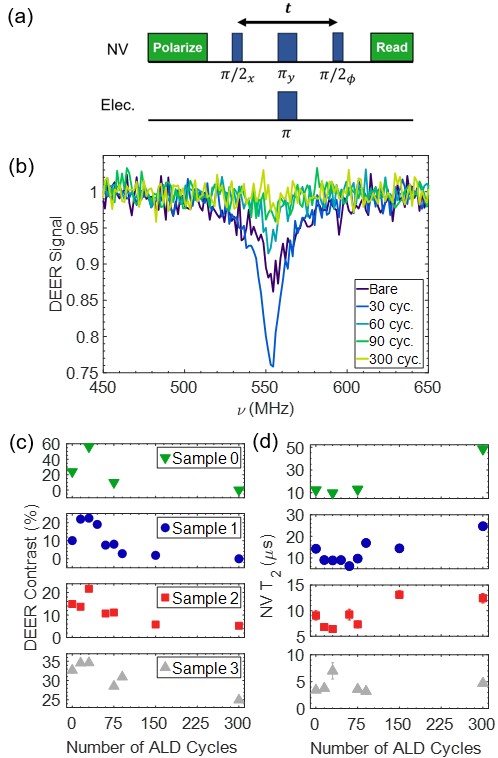}
\caption{DEER spectroscopy measurement of surface spins. 
(a) Pulse sequence for DEER measurement. The phase of the last $\pi/2$ pulse is denoted as $\phi$ to indicate the $\pm x$ phases used for variance detection normalization. 
(b) DEER measurements for Sample 1. The surface spins are probed for a total free precession time $t = 1.6~\mu s$  and a surface spin $\pi$ pulse $\sim$ 120 ns is determined from DEER Rabi measurements. (c) DEER contrast and (d) NV Hahn-echo coherence versus the number of ALD growth cycles for all samples. Single-NV data (Sample~0) were acquired at a magnetic field \(B\approx 400~\mathrm{G}\) with \(t=4\,\mu\mathrm{s}\) for the DEER measurement, whereas ensemble data (Samples~1--3) were acquired at \(B\approx 200~\mathrm{G}\) with \(t=1.6\,\mu\mathrm{s}\).}
\label{fig:DEER}
\end{figure}

At the heart of our investigation lies an NV sensing version of double electron-electron resonance (DEER) spectroscopy that has been adapted for the detection of surface dark spins \cite{NewJPhys.13.055004, PhysRevB.86.195422, PhysRevLett.113.197601}. 
In short, the DEER sequence consists of a Hahn-echo applied on the NV center(s) and a synchronized microwave $\pi$-pulse probing the resonance frequency of nearby surface spins (Fig. \ref{fig:DEER}(a)). 
If the microwave $\pi$-pulse matches the surface spin resonance, the DEER sequence translates the surface spin-NV coupling into an observable fluorescence signal and provides a qualitative measure for the coupling. 

We measure DEER at $\sim \,$200 G (\textit{B} $\parallel [111]$) for Sample 1 in Fig.~\ref{fig:DEER}(b) using a custom-built microscope (Appendix \ref{Experimental Setup}). 
We find that the observed resonance contrast strongly depends on the number of ALD cycles and, therefore, the \ch{TiO2} thickness. 
Specifically, we observe an increase in DEER signal contrast from $10.08 \pm 0.53$\% for a bare diamond surface to $22.50 \pm 0.57$\% for a diamond exposed to 30 cycles of ALD growth.
Although this amounts to a more than two-fold increase in DEER signal, the deposition of a thicker layer \ch{TiO_2} does not further increase the signal magnitude but rather leads to a reduction. 
For 300 cycles of ALD growth, corresponding to a \ch{TiO_2} layer approximately 14 nm thick (Fig.~\ref{fig:MatCharFig}(b)), the DEER signal falls below the detection limit of our measurements. 
The initial increase in DEER signal as a function of number of ALD cycles corresponds to the first phase of island growth when the surface is artificially roughened. The subsequent decrease in DEER signal occurs when the film growth becomes linear.
This trend holds across all four investigated samples (Fig.~\ref{fig:DEER}(c)).  
Furthermore, as the \ch{TiO2} thickness increases, we observe a decrease in photoluminescence for Samples 1-3, which we detail in Appendix \ref{NVPLappendix}.

%
As the nitrogen implantation dose increases, the corresponding DEER contrast increase from bare to 30 ALD cycles gets smaller.
%
Additionally, a finite DEER signal remains after growth of 300 ALD cycles on Samples 2 and 3 (see Appendix~\ref{P1}). 
This can be explained by the increasing number of bulk defects from lattice damage at higher implantation dosages leading to residual signals on resonance with the surface spins.
The density of substitutional nitrogen impurities (P1 centers) also increases with implantation dose. This leads to residual NV-P1 coupling and the appearance of additional resonances (Sample 2: four hyperfine \textsuperscript{15}N peaks, Sample 3: six hyperfine \textsuperscript{14}N peaks) within the frequency span of our measurements. We note that the signal-to-noise ratio in our measurements limits the number of detectable P1 resonances to two for Sample 2. 
We derive the P1 resonance conditions and label them in our DEER results for Samples 2 and 3 in Appendix~\ref{P1} to emphasize their spectral separation from the surface spin resonance. 
Importantly, this separation of $\geq 10$ MHz indicates that the presence of NV-P1 coupling does not measurably affect the surface spin experiments outlined in this work.
Additionally, we use this information to independently confirm the g-factor of the measured dark spins to be $g = 2.0067(21)$ (see Appendix \ref{g-factor}), which is in good agreement with previous reported results for surface spins on bulk diamond \cite{NewJPhys.13.055004} and high-field EPR measurements on diamond nanocrystals \cite{10.1063/1.5085351}.
%

Following these results, we assess the impact of the \ch{TiO_2} coatings on NV coherence ($T_2$). The results are shown in Fig.~\ref{fig:DEER}(d). 
For Sample 0, coating with 300 ALD cycles enhances the NV Hahn-echo $T_2$ by $3.8\pm0.10$ fold; time traces from 10 individual NV centers are treated in an ensemble manner (summed and fit to a single-exponential), and the individual $T_2$ results and enhancement versus depth are given in Appendix~\ref{NVT2appendix}.

By comparison, Sample 1 shows a smaller $T_2$ enhancement by a factor of $1.73\pm0.07$ for the same number of cycles, while Samples 2 and 3 show a more modest increase by a factor of $1.37\pm0.17$ and $1.37\pm0.07$.
The diminished NV $T_2$ increase for Samples 2 and 3 can be understood by the presence of a remaining paramagnetic spin bath that is not completely suppressed by our surface coating for higher nitrogen implantation doses. 
Interestingly, NV $T_2$ across all samples shows a minimum between 30 to 60 ALD growth cycles, which corresponds to the regime of island growth where an increased DEER contrast is measured.
%
%

NV coherence time ($T_2$) is strongly anti-correlated with DEER contrast in Samples 0–2 (Pearson $r_p=-0.669,-0.675,-0.839$), but only weakly correlated in Sample 3 ($r_p=0.216$), as shown in Fig.~\ref{fig:DEER}(c,d). This outlier is consistent with decoherence in Sample 3 being dominated by bulk defects rather than near-surface dark spins, which couple only weakly to DEER. More generally, DEER contrast reflects not only dark spin density but also the NV--bath separation and bath dynamics during the measurement (e.g., relaxation). Crucially, electron spins that remain static over the Hahn-echo timescale produce quasi-static shifts of the NV resonance that are refocused by the Hahn-echo and therefore do not shorten the measured $T_2$, even though they remain visible in DEER. To study DEER contrast and NV $T_2$ more carefully, we next investigate the surface dark spin longitudinal relaxation $T_1$.

\subsection{Dark spin relaxation}

We turn our attention to systematically probing the dark spin relaxation ($T_1$) time. 
In principle, the dark spin can be probed by existing correlative DEER measurement sequences as those discussed in Ref. ~\cite{PRXQuantum.3.040328,de2012controlling,goldblatt2024sensing,PhysRevLett.113.197601}. 

\begin{figure}[ht]
    \centering
    \includegraphics[width=0.48\textwidth]{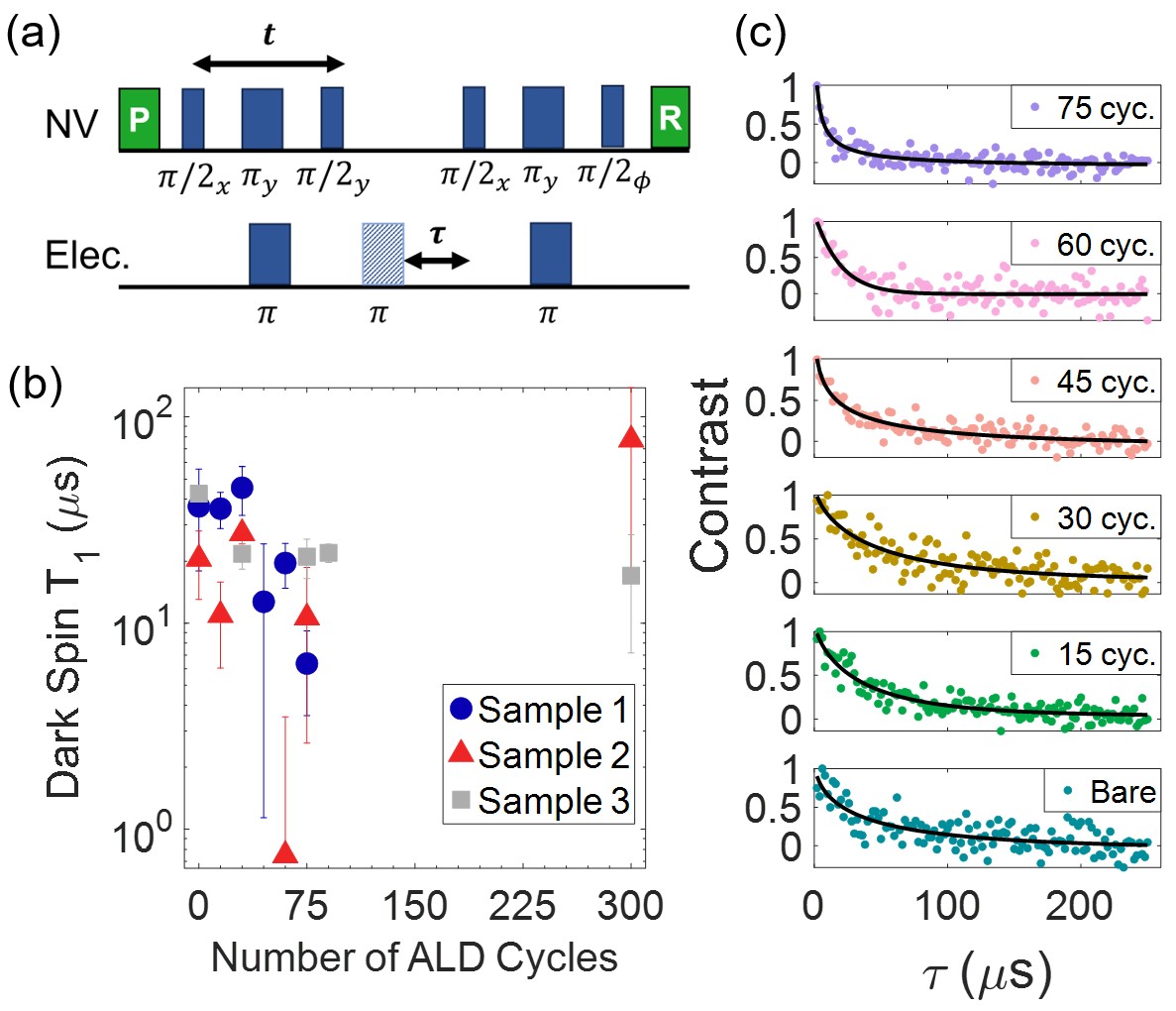}
    \caption{Dark spin relaxation characterization. (a) Pulse sequence for dark spin $T_1$ measurement including laser (green) and microwave (blue) pulses. Laser ``P" (``R") block denotes polarization (readout) pulse. (b) Dark spin $T_1$ as a function of the number of \ch{TiO_2} ALD growth cycles for the three NV ensemble diamonds. For all measurements, $t = 1.6 \, \mu s$. (c) Time traces of the $T_1$ measurement for Sample 1 up to 75 ALD cycles. Past this coating, the $T_1$ for Sample 1 is undetectable. Black curves represent the fits of the form given in Appendix \ref{SI:DarkSpinT1}.}
\label{fig:DarkT1}
\end{figure}

Here we use sequences that are NV \(T_1\)-limited~\cite{PhysRevLett.113.197601}, in contrast to protocols that are \(T_2\)-limited~\cite{PRXQuantum.3.040328,goldblatt2024sensing}. Fig.~\ref{fig:DarkT1}(a) shows a depiction of such a sequence, comprising two probe segments separated by an evolution interval. The electron spin pulse sequence measuring $T_1$ is shown in light blue. The separation between the two probe segments is adjusted according to $\tau$ in the evolution segment. If the separation between the probe segments is varied, the nuclear spin contribution encoded within the two probe segments may introduce unintended modulation~\cite{laraoui2013high}. To overcome this issue, we modify the sequence as follows: let $F^{A}_{\phi}$ be the measurement result for $T_1$,  where $A$ indicates the duration of the microwave pulse in the $T_1$ evolution segment and $\phi$ indicates the phase of the final $\frac{\pi}{2}$-pulse acting on the NV center. By taking the result $F^{\pi}_y - F^{0}_y$, the nuclear modulation will be subtracted out (see Appendix \ref{SI:DarkSpinT1} for a derivation of the sequence).
An additional reference measurement, with the last $\frac{\pi}{2}$-pulse phase shifted by 180$^o$ was also performed to cancel the common-mode noise due to NV center control. The complete sequence is given by

%
%

%
%
%
\begin{figure*}[ht]
    \centering
    \includegraphics[width=\textwidth]{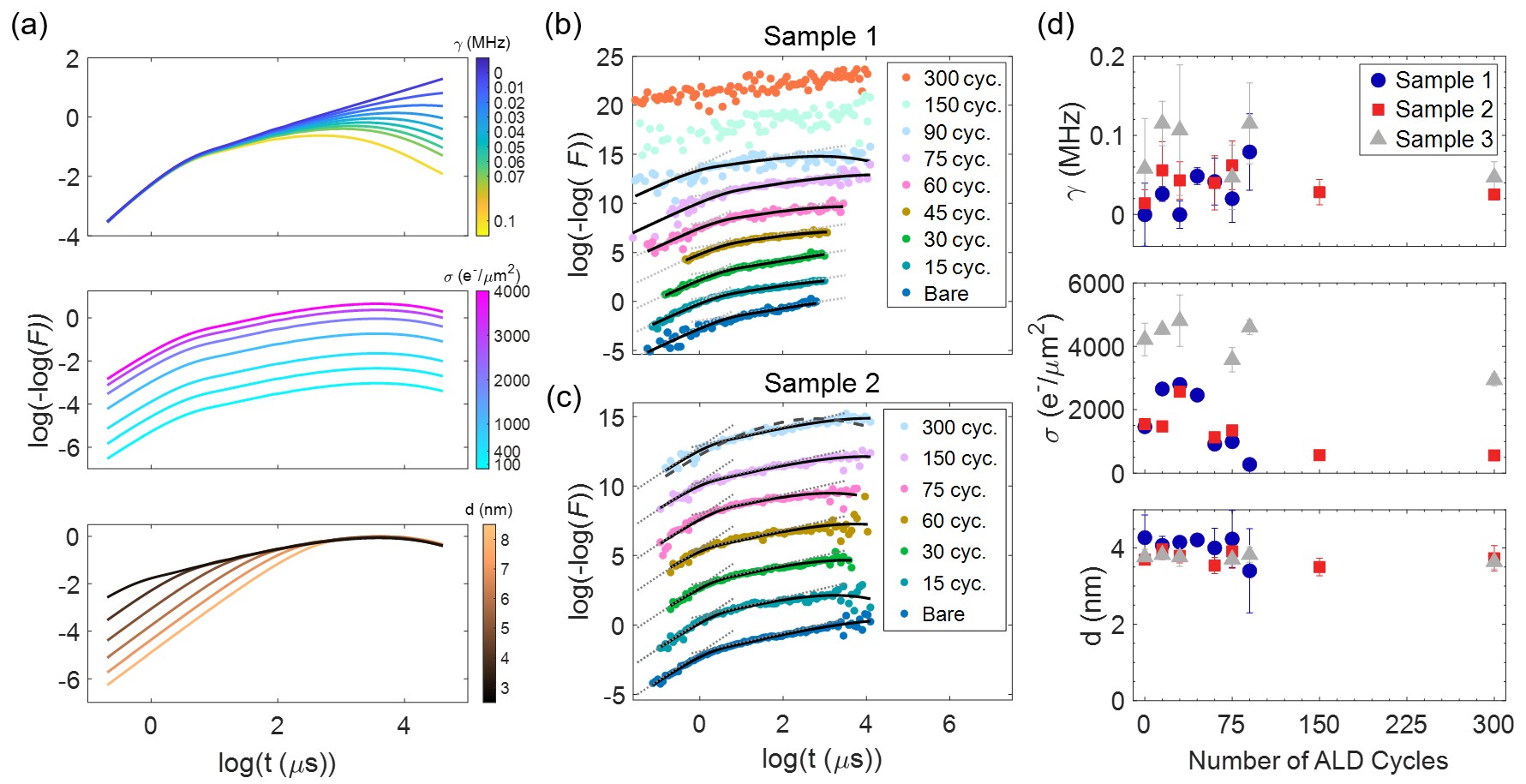}
    \caption{Model fitting for normalized DEER coherence measurement. (a) Computed $\log( -\log(\langle F(t) \rangle ))$ curve dependencies based on changing values of dark spin relaxation rate ($\gamma$, top), dark spin density ($\sigma$, middle) and NV-bath separation ($d$, bottom). (b) Sample 1 and (c) Sample 2, $\log( -\log(\langle F(t) \rangle ))$ curves with model fits. Solid black lines indicate model fit assuming a 2D layer of spins at the diamond surface which agrees well with the data.  Dark gray dashed line for Sample 2, 300 cycle data indicates model fit assuming a 3D layer of spins which poorly matches the curve profile.  Light gray dotted lines indicate lines with slope 2 and 2/3 respectively. The curves are artificially offset vertically for clarity. (d) Extracted fit parameters ($\gamma$ top, $\sigma$ middle, $d$ bottom) for the three NV ensemble diamonds.}
\label{fig:FIDModel}
\end{figure*}

\begin{equation} 
\begin{split} 
    \text{\it{S}} =& \left( F^{\pi}_y - F^{0}_y \right)
    - \left( F^{\pi}_{-y} - F^{0}_{-y} \right). 
\end{split} 
\end{equation}

%
%

We find that in the early island growth regime (i.e., less than 100 ALD cycles), the dark spin $T_1$ times decrease (Fig.~\ref{fig:DarkT1}(b)) across Samples 1-3 with the effect most prominent in Sample 1. Time traces with fits for Sample 1 are shown in Fig.~\ref{fig:DarkT1}(c) (see Appendix \ref{SI:DarkSpinT1} for details on the fit procedure). 
Beyond 75 cycles of \ch{TiO_2} growth, the $T_1$ decay is not detectable. 
Furthermore, we observe a qualitatively similar decrease in NV $T_1$ as a function of increasing \ch{TiO_2} layer thickness (Appendix~\ref{NVT1Fit}), suggesting that the \ch{TiO_2} layer induces high-frequency noise that leads to dark spin and NV relaxation. 

\subsection{Dark spin density and spatial configuration} 
The previous measurements provide valuable insights into dark spin $T_1$, but do not provide information on dark spin density. Next, we investigate the density and spatial configuration of the dark spin bath. NV dephasing caused solely by interaction with dark spins can be described by the coherence function $\langle F(t) \rangle = \exp \left(\sigma\int\int_S\{ f_{DEER}(t)-f_{Echo}(t)\} dxdy \right)$, where $\sigma$ denotes the dark spin density and the integration domain $S$ runs over the diamond surface. 
The factors $f_{DEER}$ and $f_{Echo}$ denote the NV coherence from the DEER sequence shown in Fig.~\ref{fig:DEER}(a) and a conventional Hahn-echo, respectively ~\cite{PRXQuantum.3.040328, davis2023probing}. 
Going beyond the standard treatment of a static bath, we incorporate the dark spin relaxation rate ($\gamma=\frac{1}{T_1}$), resulting in the coherence difference
%
\begin{multline}
    f_{DEER}(t) - f_{Echo}(t) \\ 
    = -e^{-\gamma t/2} \frac{2 V_{dd}(\vec{r})^2 \sin^2(t\sqrt{V_{dd}(\vec{r})^2-\gamma^2}/4)}{V_{dd}(\vec{r})^2-\gamma^2}, 
\end{multline}
%
where $ V_{dd}(\vec{r})$ denotes the dipolar interaction between the NV center and a dark spin at position $\vec{r}$ (see Appendix \ref{DEER FID Derivation} for detailed derivation). 
Our model for $\langle F\rangle$ predicts distinct functional dependences on the dark-spin relaxation rate $\gamma$, the bath-spin areal density $\sigma$, and the NV--bath separation $d$. The solutions of $\langle F(t)\rangle$ for a range of typical values of these three parameters are shown in Fig.~\ref{fig:FIDModel}(a). 

We measure $\langle F(t) \rangle$ by sweeping $t$ in the DEER sequence (Fig.~\ref{fig:DEER}(a)) and conventional Hahn-echo sequence, and then fitting the result $\log( -\log(\langle F(t) \rangle ))$ to our model.
The fits for Samples 1 and 2 are shown in Fig.~\ref{fig:FIDModel}(b,c) respectively as solid black lines. 
For short times, the coherence follows a ballistic evolution as indicated by an exponent of $2$, whereas for long times, the phase evolution has a distinct diffusive character with exponent $2/3$ (dotted gray lines). This is a signature of a two-dimensional dark spin bath \cite{davis2023probing}. 
For Sample 1, the two-dimensional bath fits $\log( -\log(\langle F(t) \rangle ))$ well for coatings up to 90 ALD cycles, where a dark spin DEER signal remains detectable (see Fig.~\ref{fig:DEER}(b,c)). 
Thicker coatings show no clear dark spin resonance, and the averaged coherence $\langle F(t)\rangle$ remains close to 1. This indicates that there is negligible impact of dark electron spins on NV coherence. When computing $\log(-\log(\langle F(t)\rangle))$, the small deviations from 1 are strongly amplified, which accounts for the noise in the 150-cycle and 300-cycle Sample 1 data in Fig.~\ref{fig:FIDModel}(b). In this regime, the model effectively breaks down because the surface dark-spin contribution is too small to be reliably extracted.
By contrast, there is never full suppression of the dark spin resonance for Sample 2 (see Fig.~\ref{fig:DEER}(c)). As such, the two-dimensional noise model fits well for all coatings measured. 
For reference, a three-dimensional dark spin noise model is not suitable for Sample 2, demonstrated by the poor fit for the 300 ALD cycle data (dashed black line in Fig.~\ref{fig:FIDModel}(c)). 
We note that for very long evolution times, relaxation of the dark spin bath must be considered, as indicated by a deviation of our model from the exponent of $2/3$. 

Having developed a model for our experimental data, we can extract $\gamma$, $\sigma$, and $d$ for our various ensemble samples, shown in Fig.~\ref{fig:FIDModel}(d) (see Appendix \ref{DEER FID fitting} for fitting procedure). 
We find that $\gamma$ increases with increasing \ch{TiO_2} thickness, as corroborated by the independent surface spin $T_1$ measurements in Fig.~\ref{fig:DarkT1}(b). 
Similarly, $\sigma$ increases during the initial phase of island growth and then decreases with increasing ALD cycles as the islands converge and the growth rate becomes linear. This qualitatively mirrors the DEER contrast behavior observed in Fig.~\ref{fig:DEER}(c). 
This suggests that \ch{TiO_2} is an effective method for suppressing parasitic dark spins on the surface, and the NV $T_2$ improvement in Fig.~\ref{fig:DEER}(d) further corroborates this finding.
Lastly, $d$ remains constant as a function of the number of \ch{TiO_2} ALD cyles, indicating that the \ch{TiO_2} film itself does not host additional dark spins --- at least not dark spins that are detectable by our DEER readout.

\subsection{Atomistic models of diamond-T{\lowercase{i}}O$_2$ heterostructures} \label{Atomistic DFT models}
Finally, we investigate the electronic structure of the diamond-\ch{TiO2} interface by formulating atomistic models based on structural information derived from our XPS measurements (Fig.~\ref{fig:MatCharFig}(d,e)). 
We employ density functional theory (DFT) to calculate the band alignment at the interface and assess the effectiveness of \ch{TiO2} as a passivation layer in mitigating surface dark spins or dangling bonds.
%

\begin{figure}[!ht]
    \centering
    \includegraphics[width=0.46\textwidth]{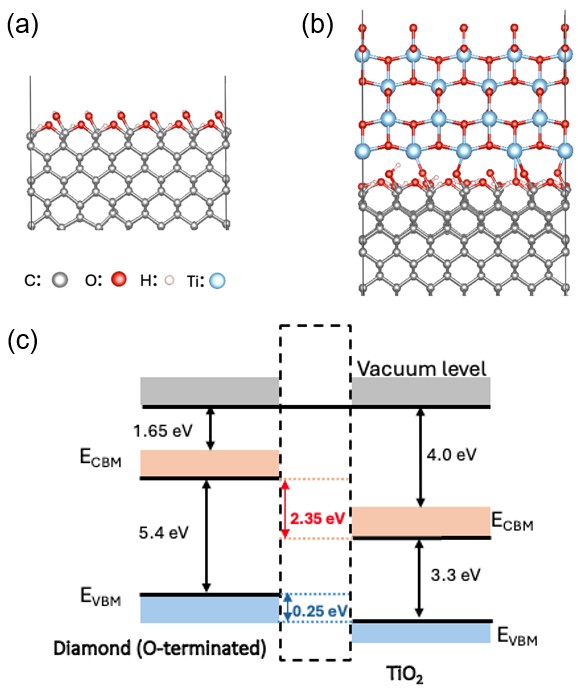}
    \caption{(a) Surface model of diamond C (100) terminated with with C-O-C, C-H, C-O, and C-OH groups. 
    (b). Heterostructure of C(100):TiO$_2$ (101).
    (c) Band alignment of diamond and TiO$_2$ from DFT. The $E_{CBM}$ (peach color) and $E_{VBM}$ (blue color) represents the conduction and valence band edges respectively. The Conduction band offset (CBO) is 2.35 eV, and the Valence band offset (VBO) is 0.25 eV.}
    \label{fig:DFT}
\end{figure}

The prototype heterostructure model consists of a (100) diamond slab, whose surface is terminated with various chemical functional groups, including carbonyl (\ce{C=O}), ether (\ce{C-O-C}), and hydroxyl (\ce{C-OH}) configurations (Fig.~\ref{fig:DFT}(a)). This is interfaced with a Ti-rich anatase TiO$_2$ slab oriented along the (101) crystallographic plane. The diamond slab has lateral dimensions of approximately \(15 \, \text{\AA} \times 15 \, \text{\AA}\)  and a thickness of \(12 \, \text{\AA}\) while the TiO$_2$ slab has the same lateral dimensions and a thickness of \(13 \, \text{\AA}\). A vacuum spacing of \(20 \, \text{\AA}\) was included along the surface normal to prevent spurious interaction between periodic images of the heterostructure. Our first-principles calculations were performed using SG15 Optimized Norm-Conserving Vanderbilt (ONCV) pseudopotentials \cite{sg15} and plane-wave basis sets, with a kinetic energy cut-off of 80 Ry, as implemented in the Quantum Espresso package \cite{giannozzi2009quantum}. The  supercell Brillouin  zones were sampled with only the $\Gamma$ point. Atomic configurations  were optimized using the gradient-corrected Perdew-Burke-Ernzerhof (PBE) functional \cite{perdew1996generalized} while electronic structure calculation were performed using the screened hybrid HSE functional \cite{hybrid}, which is known to improve the accuracy in predicting band gaps, over semi-local functionals. 

 Our results show that the initial deposition of TiO$_2$ introduces a high density of undercoordinated Ti atoms at the interface, driving significant structural rearrangements upon geometric relaxation. As shown in Fig.~\ref{fig:DFT}(b), these unsaturated Ti atoms preferentially interact with functional groups of the diamond surface, particularly \ce{C-OH} and C-O. This effectively reduces their concentration compared to the configurations existing prior to geometric relaxation. This suggests that Ti atoms can readily bond with oxygen species exposed at the surface, stabilizing the interfacial structure through the formation of \ce{Ti-O-C} linkages and facilitating the reduction of sp$^3$ carbon \cite{Dsurface}. In contrast, the ether \ce{C-O-C} groups remain largely intact, indicating their relative stability against Ti-mediated reconfiguration. 
The preservation of these groups and the selective bonding of Ti atoms highlight the role of interfacial chemistry in dictating charge transfer and the structural stability in the diamond/TiO$_2$ heterostructures prepared experimentally,  as also seen in the XPS results. Although our model is based on crystalline anatase, the insights obtained may ultimately inform the design and understanding of amorphous TiO$_2$ interfaces formed under realistic ALD conditions.

Next, we study the band alignment at the diamond-\ch{TiO2} interface using the bulk plus lineup method \cite{Martin}, which determines band offsets by referencing the bulk band edges of the two materials to the electrostatic potential difference across the interface. To align energy levels with respect to the vacuum level, we constructed isolated slab models of diamond and TiO$_2$ with the same in-plane dimensions as the heterostructure. The diamond slab, oriented along (100), had a thickness of \(12 \, \text{\AA}\), while the TiO$_2$ (101) slab was \(13 \, \text{\AA}\). In both cases, a vacuum region of \(20 \, \text{\AA}\) was added along the surface normal to eliminate interactions between periodic images.
 
%
Our simulation of the band alignment is depicted in Fig.~\ref{fig:DFT}(c). The calculation reveals a conduction band offset (CBO) of 2.35 eV and a valence band offset (VBO) of 0.25 eV. Importantly, the type of CBO facilitates electron transfer from the diamond to the TiO$_2$, while the relatively small VBO suggests the presence of minimally confined holes.  The band offsets found here suggest that the deposition of \ch{TiO2} on diamond may facilitate the transfer of electrons, which act as dark spins, into the oxide, under conditions where the diamond surface is fully passivated by \ch{TiO2}, as shown  in Fig.~\ref{fig:FIDModel}(d).

%
%

\section{Conclusion} 
In conclusion, we have introduced \ch{TiO2}-diamond heterostructures as an efficient means to engineer surface dark spin densities. 
With a comprehensive noise model that captures the effects of a relaxing dark spin bath on NV coherence, we accurately quantified dark spin densities and found the observed dark spin density conforms to a non-uniform island growth during the ALD process. 
%
%
%
%
Although our X-ray surface characterization indicates a link between ALD nucleation and dark spin density, the atomistic mechanism, and thus the origin of the surface dark spins, remains to be clarified. 
Furthermore, while our \ch{TiO2} passivation layer fully suppresses the dark spin signal at the diamond interface under ambient conditions, future studies should investigate whether chemically reactive environments can induce new interfacial spins at the \ch{TiO2}-solution boundary and how such effects can be mitigated for specific sensing applications.
Beyond ALD, TiO$_2$ can be deposited by thermal evaporation, RF magnetron sputtering, or chemical vapor deposition; however, many of these methods require relatively high temperatures ($\sim300~^{\circ}\mathrm{C}$) that may modify shallow-NV properties.

We also demonstrate through our model the two-dimensional nature of the dark spin bath, lending our technique to studies of interactions in two-dimensional spin ensembles ~\cite{rezai2025probing}. By tuning the coating parameters such that the density is approximately $20,000$~$\mu$m$^{-2}$ one will be able to enter the dipole interaction regime given the coherence of the electron spin is 200 ns (Appendix~\ref{appendix:darkspinT2}). 
In addition, controlling dark spin density and proximity to target molecules has the potential to benefit their role as reporter spins in biological sensing applications ~\cite{schaffry2011proposed,PhysRevLett.113.197601} or achieve nuclear spin hyperpolarization for better NMR sensitivity~\cite{bucher2020hyperpolarization,eills2023spin}.

%
%
Our \ch{TiO_2}-diamond heterostructures are directly compatible with existing functionalization techniques that enable immobilization of intact biomolecules on diamond surfaces~\cite{PNAS.119.8}. 
This functionalization combined with reduced dark spin density enables high-sensitivity EPR spectroscopy of intact biomolecules with practically achievable grafting densities on the order of 500 $\mu m^{-2}$. 
In such a scenario, we can expect an SNR of 1 after an estimated 15 minutes of integration time (see Appendix~\ref{EPR Sensitivity Cal}). 

%
%
%

Our methods can also be employed to passivate the surfaces of a variety of diamond structures, including probes for scanning NV microscopy. In these devices NV centers are proximal to surface damages and as such could see a dramatic benefit from our surface treatment. This extends to other nanofabricated structures, which is particularly promising given the significant progress made recently in interfacing NV centers with nanophotonic devices \cite{Addhya2024}.    
Our technique could further be applied to other qubit platforms where surface noise is a limiting factor, such as in rare earth-doped materials \cite{bartholomew2017optical,liu2020defect} and superconducting qubits \cite{bal2024systematic,chang2025eliminating}. 

\section*{ACKNOWLEDGMENTS}

The authors thank Prof. Ania Jayich, Prof. Nathalie de Leon, Prof. Shimon Kolkowitz, Dr. Yizhong Huang, Dr. Ruoming Peng, Uri Zvi and, Tian-Xing Zheng for helpful discussion related to the physics of surface spins, diamond surfaces, and possible applications. We acknowledge the use of the Pritzker Nanofabrication Facility at the University of Chicago. 
J. N., G. G., E.V. and S.W. acknowledge support by NSF QuBBE QLCI (NSF OMA-2121044) for the investigation of NV-enabled EPR bio-sensing approaches. 
NV-enabled EPR spectroscopy of surface dark spins by E.V. was supported by US DOE, Basic Energy Sciences, Division of Chemical Sciences, Geosciences, and Biosciences, through ANL under Contract No. DE-AC02-06CH11357.
Work by X.Y. on dark spin dynamics and material characterization by S.W. was supported by Q-NEXT (Grant No. DOE 1F-60579). 
M.X. was supported by NSF OMA-1936118.	
Work by J.C.J., N.D., and A.B.F.M. including atomic layer deposition of TiO2, ellipsometric measurement and fitting, XPS interpretation, and scientific guidance was supported by the U.S. Department of Energy, Office of Science, Basic Energy Sciences, Materials Science and Engineering Division. 

\appendix
\setcounter{section}{0}

\counterwithin{equation}{section} 



\renewcommand{\thesection}{\Alph{section}}
\renewcommand{\thesubsection}{\thesection.\roman{subsection}}

\makeatletter
\renewcommand{\p@subsection}{}
\makeatother

\section{Diamond Sample Processing} \label{Diamond Sample Processing}
All diamond samples measured in this work were single crystal samples grown by chemical vapor deposition purchased from Element Six. The near surface NV centers were introduced to the diamond substrates following an established procedure \cite{PhysRevX.9.031052}. The diamond samples are first etched using \ch{Ar}/\ch{Cl_2} followed by \ch{O_2} plasmas to remove the top few microns of polishing damage. The diamonds are then subjected to a vacuum anneal at $1200^{\circ}$C to intentionally graphitize the surface and remove any underlying etching damage. A triacid cleaning (1:1:1, \ch{HNO_3}:\ch{HClO_4}:\ch{H_2SO_4}) is performed to remove the graphitic layer before samples are sent for ion implantation at the desired dosages. Following ion implantation, the diamond samples are once again annealed at $800^{\circ}$C to activate the NV centers.

The diamond samples measured for ex situ variable angle spectroscopic ellipsometry and XPS are 3 identical samples implanted with $5 \times 10^{8}$~cm$^{-2}$ at 3~keV. The surfaces of each sample were processed in the same manner as Samples 0-3 measured in the main text, as outlined above. 

\section{Simplified Island Nucleation Equation} \label{INM}
The following island nucleation model is used to estimate the \ch{TiO_2} nucleation density on the diamond surface in Fig. \ref{fig:MatCharFig}. The model is a modified version adapted from Nilsen et al. \cite{Nilsen} where the unit cell is simplified to a disk from a hexagon. In equation \eqref{INM_thickness} $\mu$ is defined to be the mean thickness of the film and $N_d$ is the nucleation density $(0.047 \pm 0.017 \mathrm{nm^{-2}})$. The radius of each island assuming steady state growth of $g=0.04667 \pm 0.00098 \mathrm{nm/cycle}$ for $x$ number of cycles is given by equation \eqref{INM_growthrate}. There are two stages to island growth: the first is when the growth in unconstrained due to spare nucleation sites, and the resulting islands grow uniformly in all available directions, resulting in a hemisphere. This occurs up until the radius of the islands exceeds $R_{\mathrm{cov}}$, the radius of unit cell defined by equation \eqref{INM_disk}, which in our case is $\sim 2.6\pm0.5\mathrm{nm}$. At this point($r>R_{\mathrm{cov}}$) the growth becomes constrained by neighboring islands. For the best fitting results, both $N_d$ and $g$ are left as fit parameters, resulting in an $R^2=0.99264$.

\begin{equation}\label{INM_thickness} 
    \mu =
    \begin{cases}
     \qquad\qquad\quad \frac{2}{3}N_d\pi r^3  &r \leq R_{\mathrm{cov}}\\
    \\
    \: N_d \! \Big( \! \pi R^2_{\mathrm{cov}} \! \sqrt{r^2-R^2_{\mathrm{cov}}} \!+\! \frac{\pi}{6}\Big[3R^2_{\mathrm{cov}}+ \\
    \: \big(\!r \!-\! \sqrt{r^2 \!-\!R^2_{\mathrm{cov}}}\big)\!^2\Big]\!\big(r\!-\!\sqrt{r^2\!-\!R^2_{\mathrm{cov}}}\big)\!\Big) \qquad\ &r>R_{\mathrm{cov}}\\
    \end{cases}
    \\
\end{equation}
\begin{equation}\label{INM_growthrate}
r = gx
\end{equation}
\begin{equation}\label{INM_disk} 
    R_{\mathrm{cov}}= \sqrt{\frac{1}{\pi N_d}}
\end{equation}

\section{Solvent Stability and Biochemical Functionalization of \ch{TiO2} Coatings} \label{TiO2Stability}
\begin{figure}[ht]
    \centering
    \includegraphics[width=\linewidth]{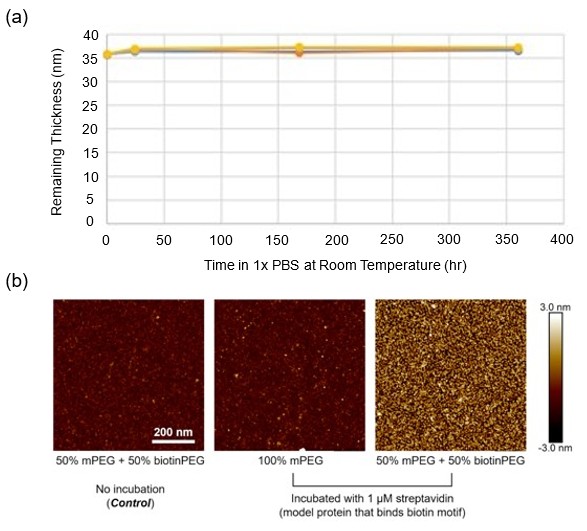}
    \caption{(a) Dissolution of \ch{TiO2} film on diamond incubated in 1X PBS over a two-week period. The remaining thickness of the coating layer at four different sites (orange, yellow, blue, and purple curves) were tracked. (b) Functionalization of Si wafers coated with 15 nm \ch{TiO2}. Silanzed surfaces underwent subsequent PEGylation with either a 50/50 mixture of biotin-containing and methyl-terminated PEG, or 100\% methyl-terminated PEG. Incubation of 1 $\mu$M streptavidin with the biotinPEGylated surface demonstrates successful immobilization of a near-monolayer of streptavidin molecules. Samples were baked at \SI{95}{\degree C} for 30 seconds prior to AFM measurement.}
    \label{figSI:TiO2dissolution}
\end{figure}

We measured the dissolution rate of our coated \ch{TiO2} film in 1X phosphate buffered saline (PBS) at room temperature by tracking its remaining thickness over time using AFM. We lithographically patterned a 35-nm-thick \ch{TiO2} film on a single-crystal diamond substrate surface. We performed AFM using a Bruker Dimension Icon instrument with SCANASYST-AIR tips to image regions on the pattern edges, from which the step height (i.e. \ch{TiO2} thickness) was extracted. As shown in Fig. \ref{figSI:TiO2dissolution}(a), four different sites of the \ch{TiO2} layer were tracked and the thicknesses remained unchanged for at least two weeks, demonstrating excellent chemical robustness under physiological conditions. AFM measurements were performed at 0, 24, 168, and 336 h. Prior to each measurement, the sample was taken out from PBS and rinsed with water, followed by drying with nitrogen gas. A slight increase in the average thickness was observed in the first 24 h but it remained stable for the remaining duration of the experiment. This stability contrasts sharply with previously reported \ch{Al2O3} coatings \cite{PNAS.119.8}, which exhibit a dissolution rate of approximately 0.74 nm/day under identical conditions.

To assess the biocompatibility of the \ch{TiO2} surface functionalization, 15-nm-thick \ch{TiO2} films were deposited on atomically flat silicon wafers. Following our previously established procedure in \cite{PNAS.119.8}, we silanized the \ch{TiO2} surface and subsequently PEGylated it using mixed polyethylene glycol (PEG) species with varying terminal groups (Fig.~\ref{figSI:TiO2dissolution}(b)). The surfaces modified with 100\% mPEG, which terminates in a methyl group, showed negligible protein adsorption on AFM after incubation with 1~\textmu M streptavidin ($\sim$5 nm in diameter) for 5 minutes and subsequent rinsing with copious amounts of water. This is indicative of effective antifouling behavior. In contrast, surfaces modified with 50\% biotin-PEG exhibited dense and uniform streptavidin binding, forming nearly close-packed protein monolayers. This controlled functionalization capability allows precise tuning of surface chemistry for specific applications, such as targeted protein immobilization in nanoscale NV-EPR or NV-NMR experiments.


\section{XPS analysis and additional data} \label{XPS}
All XPS measurements were performed using the ThermoFisher Scientific NEXSA G2 Surface Analysis System at the Keck-II facility of Northwestern University’s NUANCE Center. The measured samples were prepared following the procedure outlined in Appendix \ref{Diamond Sample Processing} to best approximate the surface conditions of the samples used for dark spin spectroscopy in the main text. All XPS spectra were fit using CasaXPS. A Shirley background is subtracted before Voigt-like functions are used to fit the remaining spectral line shape. Unless otherwise noted, the GL(30) line shape, a product of Gaussian and Lorenztian functions, was used for all peaks. 

\subsection{Photoelectron Escape Probability}

XPS is a highly surface sensitive measurement. Only the photoelectrons that do not undergo inelastic collision contribute to the peaks observed in the XPS spectra. All other photoelectrons contribute to the spectral background or otherwise do not reach the analyzer \cite{introXPS}. As such, it is important to consider the photoelectron escape probability of the material given by, $P(\mathrm{z},\lambda)=e^{\frac{-\mathrm{z}}{\lambda}}$, where z is the depth of the photoelectron in a material with inelastic mean free path (IMFP), $\lambda$. This can be critical in cases such as the present work where the samples have unique geometries and there is a particular interest in the interface between the two materials. 

Using conservative estimates for \ch{TiO2} \cite{Fuentes2002} the IMFP for carbon 1s photoelectrons is roughly 1.8nm and for oxygen 1s photoelectrons is 1.5 nm. These values are similar to established values for diamond (1.9nm for carbon 1s and 1.6nm for oxygen 1s)\cite{Tanuma2010}. It can be easily seen that with these estimates for the IMFP the photoelectron escape probability corresponds well with the changes we observe in our XPS spectra with increasing ALD cycles (Fig. \ref{fig:MatCharFig}). The peak area for photoelectrons coming from the diamond ($\mathrm{sp}^3$) reduces as \ch{TiO2} islands converge and the thickness of the film grows. And by 300 cycles when the expected \ch{TiO2} film is nearly 14nm thick, there is a negligible probability for photoelectrons to escape from the diamond and as such no discernible peaks coming from the diamond are seen in the carbon 1s or oxygen 1s spectra. 

\begin{figure}
    \centering
    \includegraphics[width=\linewidth]{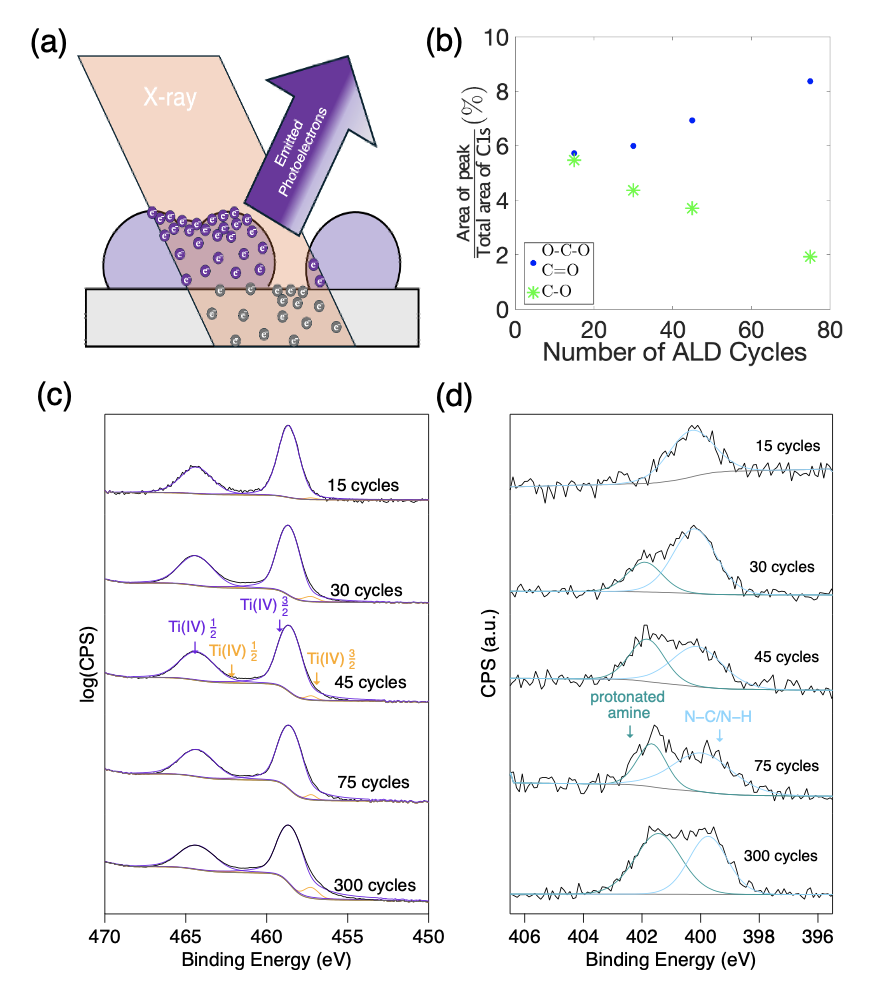}
    \caption{(a) Illustration of Photoelectron Escape Probability in XPS. (b) Comparison of peak areas in carbon 1s spectra. The area for each peak is normalized by the total area fit. (c) XPS titanium 2p spectra for varying number of ALD cycles. (d) XPS nitrogen 1s spectra for varying number of ALD cycles}
    \label{fig:XPSsupp}
\end{figure}

\subsection{Carbon 1s Spectra}

To elucidate the changes in the surface functional groups, in Fig. \ref{fig:XPSsupp}(b) the peak areas for the singly and doubly bonded carbon species are normalized by the total carbon 1s peak area. It shows a clear decrease in the percentage of singly bonded carbon and a corresponding increase in the percentage of doubly bonded carbon. Interestingly, the total percentage of signal coming from functional groups on the diamond surface remains more or less constant. This could be due in part to some peak misfitting but is also indicative that the surface chemistry is shifting from hydroxide groups to predominantly ketone or ether species as the \ch{TiO2} film grows. The atomistic model proposed in Section \ref{Atomistic DFT models} reflects these changes in surface chemistry.

\begin{table}[h]
\caption{The fit parameters for the fitted spectra shown in Figure \ref{fig:MatCharFig}(d) are shown below. Peak spacings were constrained according to experimental values. \cite{Baldwin2014}.  For fitting the adventitious carbon peaks in the 300 cycle data, all peaks in the spectra were constrained to have the same FWHM. }
\begin{tabular}{|b{3.5em}|m{3em}|m{4em}|m{4em}|m{4em}|m{3em}|}
\hline
C1s Species & ALD Cycles & Binding Energy (eV) & FWHM & Area    & \% Conc. \\
\hline\hline
$\mathrm{sp^3}$         & 15         & 285.86                         & 0.79 & 646239  & 76.88            \\
            & 30         & 286.13                         & 0.78 & 302255  & 67.95            \\
            & 45         & 286.03                         & 0.78 & 38098.5 & 41.39            \\
            & 75         & 286.13                         & 0.91 & 7719    & 25.69            \\
\hline
$\mathrm{sp^2}$         & 15         & 284.86                         & 1.23 & 100215  & 11.92            \\
            & 30         & 285.13                         & 1.45 & 96489.1 & 21.69            \\
            & 45         & 285.03                         & 1.74 & 32873.6 & 41.39            \\
            & 75         & 284.93                         & 1.49 & 19239.9 & 64.01            \\
\hline
C-O         & 15         & 286.86                         & 1.03 & 46024.4 & 5.48             \\
            & 30         & 287.13                         & 0.91 & 19407.4 & 4.36             \\
            & 45         & 287.03                         & 0.89 & 2941    & 3.7              \\
            & 75         & 287.13                         & 0.69 & 580.038 & 1.93             \\
\hline
O-C-O   & 15         & 288.16                         & 1.83 & 48127.9 & 5.73             \\
C=O            & 30         & 288.43                         & 1.81 & 26676.1 & 6                \\
            & 45         & 288.33                         & 1.95 & 5508.67 & 6.94             \\
            & 75         & 288.43                         & 1.8  & 2515.56 & 8.37    
\\
\hline
\end{tabular}
\begin{tabular}{|m{4em}|m{4em}|m{4em}|b{4em}|m{3em}|}
\hline
300 cyc. C1s species & Binding Energy (eV) & FWHM & Area    & \% Conc. \\
\hline\hline
sp2           & 285.16              & 1.35 & 101277  & 80.18    \\
C-O           & 286.66              & 1.35 & 12192.1 & 9.65     \\
C=O           & 288.11              & 1.35 & 4122.51 & 3.26     \\
O-C=O         & 289.13              & 1.35 & 8715.49 & 6.9   \\
\hline
\end{tabular}
\end{table}
\subsection{Oxygen 1s Spectra}

To avoid over-fitting, a hydroxide peak is not assigned for the spectra from samples prepared with 30, 45 and 75 ALD cycles  even though there is a corresponding oxygen singly bonded to carbon peak in the carbon 1s spectra. This decision was made because the hydroxide completely overlaps with the non-lattice oxygen peak. Also, as the peak assigned to ether bonds in the carbon 1s spectra is already greatly reduced in the 30 cycle data, an even greater reduction in the hydroxide contribution to the O1s spectra would be expected. This renders any resulting peak prohibitively small for resolution from the background. Here, it is important to recall that due to the higher binding energy of oxygen 1s species, the inelastic mean free path of photoelectrons is smaller compared to carbon 1s species. As such, the oxygen 1s spectra is less sensitive to the diamond surface as the number of ALD cycles increase. 

The non-lattice oxygen peak is in reality the combination of many peaks, including but not limited to Ti(III), adventitious carbon species incorporated in the \ch{TiO_2},  and hydroxide and ketone surface groups on the diamond. For the purposes of our analysis, the most relevant information is the preservation of the ether groups and the likely removal of hydroxide groups through nucleation or desorption.  As such, fitting the non-lattice oxygen components would not be a fruitful or reliable endeavor. 
\begin{table}[h]
\caption{The fit parameters for the fitted spectra shown in Figure \ref{fig:MatCharFig} (e) are shown below and are in agreement with experimental values. \cite{Kumar2016}}
\begin{tabular}{{|b{3em}|m{6em}|m{4em}|m{4em}|m{4em}|m{3em}|}}
\hline
ALD cycles & O1s species        & Binding Energy (eV) & FWHM & Area    & \% Conc. \\
\hline\hline
15         & C-O-C              & 532.6               & 1.96 & 179805  & 57.1     \\
           & O-Ti-O             & 530.34              & 1.7  & 78550.8 & 24.94    \\
           & C-OH               & 531.6               & 1.33 & 43028.6 & 13.66    \\
           & non-lattice O & 531.95              & 1.84 & 13524.9 & 4.29     \\
           \hline
30         & O-Ti-O             & 530.32              & 1.29 & 444245  & 71.49    \\
           & non-lattice O & 531.82              & 1.64 & 119461  & 19.23    \\
           & C-O-C              & 533.01              & 1.64 & 57625   & 9.28     \\
           \hline
45         & O-Ti-O             & 530.19              & 1.3  & 233775  & 84.32    \\
           & non-lattice O & 531.69              & 1.48 & 35080.5 & 12.66    \\
           & C-O-C              & 532.84              & 1.26 & 8388.2  & 3.03     \\
           \hline
75         & O-Ti-O             & 530.16              & 1.3  & 141151  & 86.36    \\
           & non-lattice O & 531.66              & 1.25 & 16747.6 & 10.25    \\
           & C-O-C              & 532.6               & 1.21 & 5548.98 & 3.4      \\
           \hline
300        & O-Ti-O             & 530.06              & 1.25 & 673792  & 84.38    \\
           & non-lattice O & 531.56              & 2.02 & 124692  & 15.62   \\
           \hline
\end{tabular}
\end{table}
\begin{table}[htbp]
\caption{All spectra are calibrated by aligning the Ti(IV) $\frac{3}{2}$ peak to 458.7 eV and SGL(30) line shapes were used to fit the doublets.  The fitted peak spacings and area constraints are in agreement with experimental values \cite{Delegan2015, Biesinger2010}. The area constraint forces the $\mathrm{2p}\frac{1}{2}$ species to have half of the area of the corresponding $\mathrm{2p}\frac{3}{2}$ species. The calculated $\%$ Conc. is weighted accordingly.}
\begin{tabular}{|b{3em}|m{4em}|m{4em}|m{4em}|m{4em}|m{3em}|}
\hline
ALD Cycles & Ti2p species                           & Binding Energy (eV) & FWHM & Area    & \% Conc. \\
\hline \hline
15  & Ti(IV) $\frac{3}{2}$  & 458.66              & 1.4  & 49909.5 & 49.28    \\[0.5ex]
           & Ti(IV) $\frac{1}{2}$  & 464.36              & 2.21 & 24954.8 & 49.69    \\[0.5ex]
           & Ti(III) $\frac{3}{2}$ & 457.26              & 0.79 & 515.416 & 0.51     \\[0.5ex]
           & Ti(III) $\frac{1}{2}$ & 462.46              & 1.12 & 257.708 & 0.51     \\[0.5ex]
           \hline
30  & Ti(IV) $\frac{3}{2}$  & 458.7               & 1.2  & 392279  & 48.88    \\[0.5ex]
           & Ti(IV) $\frac{1}{2}$ & 464.4               & 2.13 & 196139  & 49.28    \\[0.5ex]
           & Ti(III) $\frac{3}{2}$ & 457.3               & 1.01 & 7360.07 & 0.92     \\[0.5ex]
           & Ti(III) $\frac{1}{2}$ & 462.5               & 1.74 & 3680.04 & 0.92     \\[0.5ex]
           \hline
45  & Ti(IV) $\frac{3}{2}$  & 458.66              & 1.19 & 199507  & 49.34    \\[0.5ex]
           & Ti(IV) $\frac{1}{2}$  & 464.36              & 2.12 & 99753.7 & 49.75    \\[0.5ex]
           & Ti(III) $\frac{3}{2}$ & 457.26              & 0.77 & 1838.03 & 0.45     \\[0.5ex]
           & Ti(III) $\frac{1}{2}$ & 462.46              & 1.18 & 919.015 & 0.46     \\[0.5ex]
           \hline
75  & Ti(IV) $\frac{3}{2}$  & 458.65              & 1.17 & 120246  & 49.39    \\[0.5ex]
           & Ti(IV) $\frac{1}{2}$  & 464.35              & 2.1  & 60123.1 & 49.8     \\[0.5ex]
           & Ti(III) $\frac{3}{2}$ & 457.25              & 0.68 & 982.174 & 0.6      \\[0.5ex]
           & Ti(III) $\frac{1}{2}$ & 462.45              & 0.85 & 491.087 & 0.6      \\[0.5ex]
\hline
300 & Ti(IV) $\frac{3}{2}$  & 458.68              & 1.16 & 608355  & 48.95    \\[0.5ex]
           & Ti(IV) $\frac{1}{2}$  & 464.38              & 2.11 & 304177  & 49.35    \\[0.5ex]
           & Ti(III) $\frac{3}{2}$ & 457.28              & 0.98 & 10490.8 & 0.84     \\[0.5ex]
           & Ti(III) $\frac{1}{2}$ & 462.48              & 2.09 & 5245.4  & 0.85    \\[0.5ex]
           \hline
\end{tabular}
\end{table}
\subsection{Titanium 2p Spectra}
In the titanium 2p spectra we expect the primary species to belong to fully coordinated titanium (Ti(IV)). In order to place an upper limit on the concentration of paramagnetic \ch{Ti3+} ions in the \ch{TiO2} film, we force the inclusion of the Ti(III) doublet. The resulting fit shows that the paramagnetic defect concentration has no obvious dependency on the number of ALD cycles, and is therefore not the source of changes observed to the dark spin environment. Furthermore, our fits limit the concentration to $<2\%$ of the total \ch{Ti} in the film. This is an overestimate of the true paramagnetic defect concentration because the fits do not require the inclusion of the Ti(III) doublet for convergence. Indeed, fitting the spectra with Ti(IV) doublets alone results in a reasonable fit.

\subsection{Nitrogen 1s Spectra}
Across all samples measured, the nitrogen 1s amounts to $<1\%$ of the total peak area fitted across all spectra. There is no discernible increase in the nitrogen 1s signal with an increasing number of ALD cycles. Although there is prior work observing amine groups in similar \ch{TiO2} materials with XPS, an explanation for these peaks comes from the TDMAT precursor used in our ALD recipe. The amine groups likely result from residual ligands that did not react with the water pulses, similar to the growing \ch{sp^2} carbon signal seen in the carbon 1s spectra.\cite{Dufond2020}

\section{Experimental Setup}\label{Experimental Setup}
For the ensemble NV measurements, we use a custom-built benchtop epifluorescence inverted microscope. We excite the NV ensemble using 515 nm laser excitation (Oxxius LBX-515) focused just in front of the back mounting plane of an Olympus PLN 100X Oil Immersion Objective (Thorlabs RMS100X-O) to excite an NV area of approximately 300 $\mu m ^2$. A dichroic mirror (Chroma T610lpxr) and long pass filter (Semrock BLP01-594R-25) are used on the collection path to spatially separate and filter out the excitation light from the NV fluorescence, which is detected using an avalanche photodiode (Thorlabs APD410A).  The output voltage signals from the APD are recorded using a National Instruments (NI) 9223 voltage input module installed in an NI cDAQ-9185 data acquisition chassis. 

Microwaves for NV spin state control are generated by a vector signal generator (Stanford Research Systems SG396) while microwaves for surface spin driving are programmed to an arbitrary waveform generator (Zurich Instruments HDAWG4). The two microwave sources are combined through a power combiner (Mini-Circuits ZN2PD2-63-S+) and sent through a high power amplifier (Mini-Circuits ZHL-25W-63+) before reaching the sample.  The microwaves are delivered to the diamond by a coplanar waveguide with a omega-shaped antenna and the pulse sequences are coded into a Pulse Streamer 8/2 (Swabian Instruments).  

A pair of permanent neodymium magnets (K\&J Magnetics) are positioned on either side of the diamond sample based off the design in \cite{buchernatprotocols}. The magnets are mounted in a multi-stage mount that permits alignment in a spherical coordinate system. The azimuthal and polar angle adjustment is controlled by two rotation stages (Thorlabs HDR50) and the magnetic field strength by changing the distance between the magnets on linear stages (Zaber LRT0100AL-CT3A).  The polar angle is set to $\theta = 36^{\circ}$ for the (100) diamonds studied in these experiments, allowing for alignment to one of the four NV axes.

For the measurements conducted on Sample 0, the electronic spin of individual NV centers was initialized and read out using a 520-nm green laser (Labs-Electronics, DLnsec). The NV spins were coherently manipulated by a microwave signal generator (Stanford Research Systems, SG 396) with a build-in in-phase and quadrature (IQ) modulator. The dark electron spins were coherently manipulated by a microwave source (SignalCore, SC5511A) with an external IQ modulator (Polyphase Microwave, AM0260A). The microwave pulse phase and length were controlled by an arbitrary waveform generator (Zurich Instrument, HDAWG8-ME) via IQ modulation. The modulated microwaves were combined through RF power combiner (Mini-Circuits, ZACS242-100W+) and then amplified with a high-power amplifier (Mini-Circuits, ZHL-16W-43+) and delivered via a coplanar waveguide to the diamond sample. A home-built confocal microscope was used to collect NV fluorescence, which was equipped with a dichroic beam splitter (Chroma, T610lpxr) to separate excitation and emission pathways. The emission was detected by a single-photon counter (Excelitas SPCM-AQRH-14) and processed by a time tagger (Swabian Instruments, Time Tagger 20). Confocal scanning was achieved by a piezo scanner (Mad City Lab, NANOM 350). External magnetic fields were provided by a neodymium-permanent magnet (K\&J Magnetics), which was mounted on a motorized translation stage with four degrees of freedom (Zaber technologies) for full control over the magnetic field alignment. The involved devices were synchronized and triggered by a transistor–transistor logic pulse generator (Swabian Instruments, Pulse Streamer 8/2).

\section{Dark Spin Double-Electron Resonance and Coherence Measurements} \label{DEER Measurement}
\subsection{DEER}
The NV center is manipulated by a standard Hahn-echo sequence. During the midpoint of the NV free evolution, we apply simultaneous $\pi$ pulses on the NV and dark spin, determined independently from Rabi oscillation experiments targeting each spin (Appendix~\ref{darkspinmeasurements} for detail), to ensure the NV phase accumulated is purely due to dark spins. 

The Hahn-echo sequence filters out unwanted low-frequency noise by canceling the NV phase accumulation before and after the NV $\pi$ pulse. The addition of the dark spin $\pi$ pulse ensures that only the phase accumulation of the NV center due to the dark spin is preserved and detected. Since we are interested in isolating the effects of the surface electron spins on the NV ensembles, we normalize our DEER data by a regular Hahn-echo experiment to cancel unwanted effects from \ch{^{13}C} and \ch{^{15}N} nuclear spin oscillations and decoherence from other sources. In total, the DEER pulse sequence consists of four sub-sequences: two DEER echos with a 180$^\circ$ phase shift in the final $\pi/2$ pulse and two Hahn-echos with the same final $\pi/2$ pulse phase shift for normalization (variance detection).  We note that the DEER and Hahn-echo experiments are performed alternately, one after the other, repetitively, to manage systematic errors.

As an initial test to estimate the Rabi frequency of the dark spins, we use the NV centers to mimic the surface spins. First, we adjust the magnetic field to $\sim$825 Gauss so that the resonance frequency of the NV center's $|0\rangle$ state and $|-1\rangle$ state matches the expected ESR frequency (560 MHz). By performing a Rabi measurement on the NV center at this field using the dark spin microwave source, we obtain the NV Rabi frequency which serves as a proxy for the estimated dark spin Rabi frequency at 200 Gauss. We note that due to NV-nitrogen hyperfine splitting ($\sim$2-3 MHz), this method can overestimate the actual Rabi frequency of the dark spin at lower magnetic fields.

Next, we measure the dark spin resonance using the DEER sequence, keeping the phase accumulation time of the NV center to be constant while sweeping the frequency of the dark spin $\pi$-pulses. After measuring the DEER frequency, we fix the dark spin frequency at the measured value and conduct correlated measurements to determine the Rabi frequency of the dark spin accurately (Fig.~\ref{figSI:Corr-Rabi}). Finally, the DEER frequency is measured again with precise parameters.

\begin{figure*}[!ht]
    \centering
    \includegraphics[width=\textwidth]{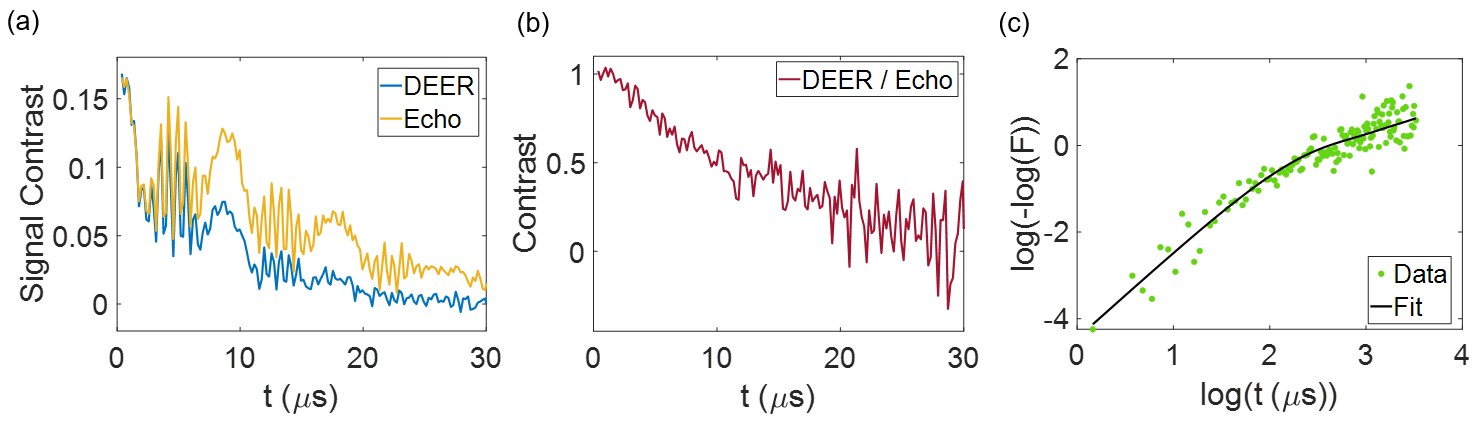}
    \caption{Sample 0 FID measurement. (a) DEER $T_2$ (blue) and Hahn-Echo $T_2$ (yellow) demonstrating \textsuperscript{15}N modulation at two frequencies. (b) FID result after dividing DEER $T_2$ by Hahn-Echo $T_2$. Division eliminates the nuclear spin oscillations appearing in the two constituent curves. (c) Double-log plot of FID curve fitted using model (black line).}
    \label{figSI:Sample0DEER}
\end{figure*}

\subsection{DEER FID Curve}
\label{DEER FID Derivation}
We consider an NV center coupled to a single electron spin via dipolar interaction. Due to the large detuning (GHz scale) between the NV resonance and the electron spin resonance compared with the interaction (kHz scale), only the term \(\sigma_z^{(v)}\sigma_z^{(e)}\) in the interaction is preserved. The other terms are energetically suppressed.  

In the rotating frame of the NV center resonance and the electron spin resonance, the free evolution of the system is governed by the following master equation:  

\begin{equation}
\begin{split}
\frac{d\rho}{dt} = &-i\left[\frac{V_{dd}}{4}\sigma_z^{(v)}\sigma_z^{(e)}, \rho\right]\\  
&+ \frac{\gamma}{2} \left( \mathcal{D}[\sigma_+^{(e)}] + \mathcal{D}[\sigma_-^{(e)}] \right) 
+ \frac{\gamma_{e_2}}{2} \mathcal{D}[\sigma_z^{(e)}],  
\end{split}
\end{equation}
where \(\mathcal{D}[A] = A \rho A^\dagger - \frac{1}{2}\{A^\dagger A, \rho\}\) is the Lindblad superoperator,  \(\gamma\) and \(\gamma_{e_2}\) are the relaxation and dephasing rates of the electron spin, respectively, $V_{dd}= \frac{\mu_0\gamma^2\hbar}{4\pi r^3}(1-3\cos^2(\theta))$ is the dipolar interaction between the NV and electron spin,  $\mu_0 = 4\pi \times 10^{-7} \, \text{H/m}$ represents the permeability of free space, $\gamma_e = 2\pi \times 28 \times 10^9 \, \text{Hz/T}$ is the gyromagnetic ratio of the electron, and $\hbar = 1.055 \times 10^{-34} \, \text{J$\cdot$s}$ denotes the reduced Planck constant.

The initial state of the system is given by \(\rho(0) = \frac{1}{4}(\sigma_0^{(v)} + \sigma_z^{(v)})\sigma_0^{(e)}\), where the NV center is polarized along the \(\sigma_z\)-axis, and the electron spin is in a maximally mixed state.  

To compute the DEER signal, the following pulse sequence is applied:  

\begin{enumerate}
    \item A \((\frac{\pi}{2})_x\) pulse is applied to the NV center.
    \item The system evolves freely for a duration of \(\frac{t}{2}\).
    \item A \((\pi)_y\) pulse is applied to the NV center and another $\pi$ pulse is applied to the electron spin.
    \item The system evolves freely for another \(\frac{t}{2}\).
    \item A final \((\frac{\pi}{2})_x\) pulse is applied to the NV center.
\end{enumerate}

The resulting system state after the pulse sequence is denoted by \(\rho(t)\). The final measurement is performed on the NV center in the \(\sigma_z^{(v)}\) basis. The DEER signal is then given by $f_{\text{DEER}} = \text{Tr}[\sigma_z^{(v)} \rho(t)]$:

\begin{equation}\label{eq:DEER}
\begin{split}
    &f_{\mathrm{DEER}} =  \\
    & \frac{|V_{dd}|e^{-\frac{\gamma}{2}t}}{\sqrt{V_{dd}^2 \!-\! \gamma^2}} \cos \!\bigg[ \!\frac{t}{2}\!\sqrt{V_{dd}^2\! -\! \gamma^2}  \!- \!\arcsec \!\Big(\! \frac{|V_{dd}|}{\!\sqrt{\!V_{dd}^2 \!-\! \gamma^2}}\! \Big) \!\bigg]
\end{split}
\end{equation}

To compute the Hahn-echo signal, in step 3, no $\pi$ pulse is applied to the electron spin. As a result, the Hahn-echo signal is given by $f_{\text{Echo}} = \text{Tr}[\sigma_z^{(v)} \rho(t)]$:

\begin{equation}\label{eq:Echo}
\begin{split}
    &f_{\mathrm{Echo}} =  \\
    & \frac{|V_{dd}|e^{-\frac{\gamma}{2}t}}{V_{dd}^2 - \gamma^2} \!\Bigg[ \!|V_{dd}|\! - \!\gamma \!\cos\! \bigg[ \!\frac{t}{2}\! \sqrt{V_{dd}^2 - \gamma^2}   \!+\! \arccos\!\Big(\frac{\gamma}{|V_{dd}|}\Big) \!\bigg] \!\Bigg].  
\end{split}
\end{equation}

In the case where \(\gamma = 0\), we recover \(f_{\text{Echo}} = 1\), indicating no effect on NV coherence.   

For multiple electron spins without interactions among them, where the initial state is a product state and the Hamiltonians of the individual electrons commute with each other, the DEER and Echo signals can be expressed as: $F_{\text{DEER}} = \prod_i f_{\text{DEER}}^{(i)}$ and $ F_{\text{Echo}} = \prod_i f_{\text{Echo}}^{(i)}
$.

If the NV spin is coupled to additional resources via the \(\sigma_z^{(v)}\) operator (for example, coupling to a nuclear spin via \(\sigma_z^{(v)}(A_{\parallel}\sigma_z^{(n)} + A_{\perp}\sigma_x^{(n)})\)), the signals are modified as: $F_{\text{DEER}} = g_{\text{other}} \prod_i f_{\text{DEER}}^{(i)}$ and $F_{\text{Echo}} = g_{\text{other}} \prod_i f_{\text{Echo}}^{(i)}
$, where \(g_{\text{other}}\) accounts for the effects on coherence due to these additional resources. Since these resources do not interact with the electron spins, their effects are identical for both the DEER and Echo signals.

\begin{figure}[!ht]
    \centering
    \includegraphics[width=0.42\textwidth]{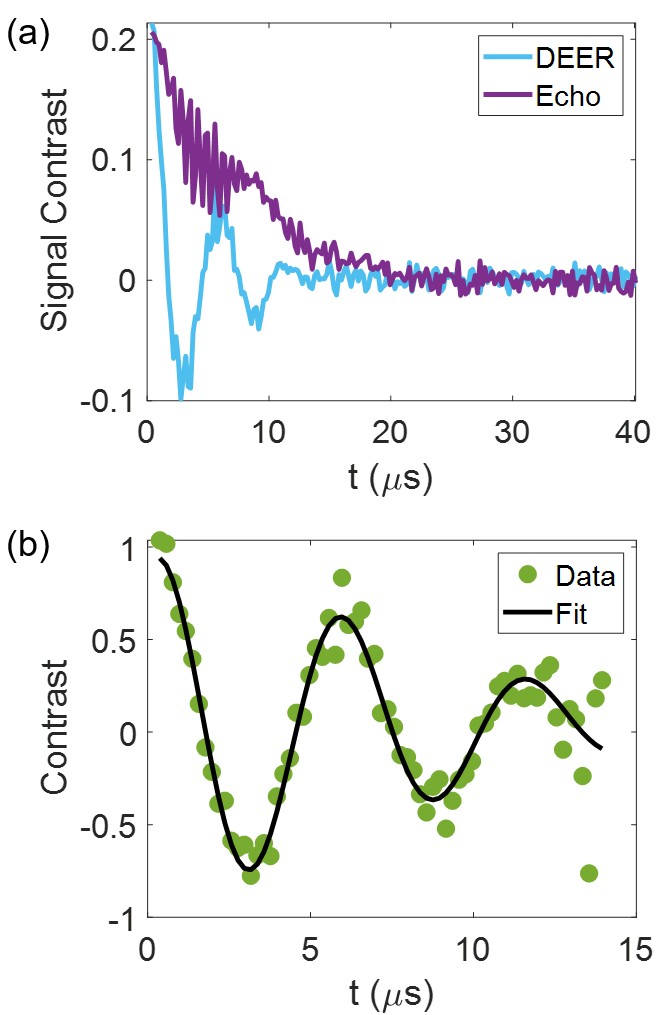}
    \caption{Coherent coupling of single NV to electron spin. (a) DEER $T_2$ (blue) and Hahn-Echo $T_2$ (purple) curves. (b) FID result from dividing DEER $T_2$ by Hahn-Echo $T_2$ showing the oscillation due to a strongly coupled electron spin.}
    \label{figSI:Sample0Coupling}
\end{figure}

Using configurational averaging ~\cite{fel1996configurational}, the averaged signals are: $\langle F_{\text{DEER}} \rangle = g_{\text{other}} e^{\sigma \int(f_{\text{DEER}} - 1) dV}$, $
\langle F_{\text{Echo}} \rangle = g_{\text{other}} e^{\sigma \int (f_{\text{Echo}} - 1)dV }$.

To cancel the effects of the additional resources, we take the ratio of the averaged signals:

\begin{equation}\label{eq:DEER-Echo-average}
\langle F \rangle = \frac{\langle F_{\text{DEER}} \rangle}{\langle F_{\text{Echo}} \rangle} = e^{\sigma\int  (f_{\text{DEER}} - f_{\text{Echo})}dV}
\end{equation}

The difference between the DEER and Echo signals is computed using Eqs. \eqref{eq:DEER} and \eqref{eq:Echo}:

\begin{equation}\label{eq:DEER-Echo}
f_{\text{DEER}}(t) \!- \!f_{\text{Echo}}(t)\! = \!-e^{-\frac{\gamma t}{2}} \frac{2 V_{dd}(\vec{r})^2 \sin^2\!\Big(\frac{t \sqrt{V_{dd}(\vec{r})^2 - \gamma^2}}{4}\Big)}{V_{dd}(\vec{r})^2 - \gamma^2}
\end{equation}

In the case where $\gamma=0$, we have $f_{\mathrm{DEER}}-f_{\mathrm{Echo}}=2\sin^2(\frac{V_{dd}t}{4})=1-\cos(V_{dd}t/2)$, which agrees with the previous result in \cite{PRXQuantum.3.040328}.

The effect of canceling out additional sources of decoherence can be demonstrated clearly with Sample 0, where individual NV centers are resolvable and can be probed. Due to slight misalignment of the external magnetic field, the DEER and Hahn-Echo sequences show nuclear spin modulation due to \ch{^{15}N}
(Fig.~\ref{figSI:Sample0DEER}(a)).This contribution to $g_{other}$ can be eliminated through dividing $F_{\mathrm{DEER}}$ by $F_{\mathrm{Echo}}$ and the result is shown in Fig.~\ref{figSI:Sample0DEER}(b). In some cases, an NV center can demonstrate strong coupling to a proximal surface electron spin (Fig. \ref{figSI:Sample0Coupling}). The oscillation curve was fitted with $Ae^{-(\frac{t}{T_2})^n}\cos(2\pi ft+\phi)+C$, where $f = 175.2 \pm 4.7$kHz  represents the coupling strength between electron spin and NV center, translating to an average separation on the order of $7$ nm. This oscillation disappears after acid cleaning the sample, indicating that these spins are mobile and reside on the diamond surface. For the non-oscillation case in Fig. \ref{figSI:Sample0DEER}(c), by using Eqs. \eqref{eq:DEER-Echo-average} and \eqref{eq:DEER-Echo}, the fit estimates the NV depth to be $7.2\pm 0.65$nm, which overlap with the value obtained via proton NMR measurement $8.45\pm 1$nm~\cite{pham2016nmr}.

\subsection{DEER Coherence Fitting}\label{DEER FID fitting}

The Free Induction Decay (FID) curve is modeled as $F(\sigma, \gamma, d, t) = e^{\sigma W(\gamma, d, t)}$, where \(W(\gamma, d, t)\) is defined as the integral $W(\gamma, d, t) = \int (f_{\mathrm{DEER}} - f_{\mathrm{Echo}}) \, dV$. We apply a double-logarithmic transformation to reveal the transition in the stretch factor~\cite{PRXQuantum.3.040328, davis2023probing}:

\begin{equation}
F_p(\sigma, \gamma, d, t) = \log\left(\!-\log(F)\right) = \log(\sigma) + \log\left(-W(\gamma, d, t)\right).    
\end{equation}

\noindent The goal is to estimate \(\hat{\sigma}\), \(\hat{\gamma}\), and \(\hat{d}\) by minimizing the cost function:

\begin{equation}
\sum_{i=1}^N|\Delta F_p(t_i)|=\sum_{i=1}^{N} \left| F_p'(t_i) - F_p(\sigma, \gamma, d, t_i) \right|^2,
\end{equation}
where \(F_p'(t_i)\) represents experimental observations. Directly solving this optimization problem is challenging due to the highly non-linear nature of \(W(\gamma, d, t)\), which lacks an analytical solution and requires computationally expensive numerical evaluation. A brute-force search over three parameters (\(\sigma, \gamma, d\)) is thus impractical.

To simplify, note that for known \(\gamma\) and \(d\), \(\sigma\) can be determined at each time \(t_i\) as: $\sigma_i = \frac{\log[F(t_i)]}{W(\gamma, d, t_i)}$. The optimal estimate for \(\sigma\) is then given by: \begin{equation}
\hat{\sigma} = \frac{1}{N} \sum_{i=1}^{N} \frac{\log[F'(t_i)]}{W(\gamma, d, t_i)}.    
\end{equation}

\noindent This approach reduces the original three-parameter optimization problem to a two-parameter search for \(\gamma\) and \(d\). Once \(\hat{\sigma}\), \(\hat{\gamma}\), and \(\hat{d}\) are determined, their uncertainties can be computed using error propagation.


The error in \(F(t_i)\) is expressed as:

\begin{equation}
    \Delta F_p(t_i)^2\!=\!\left(\frac{\partial F_p}{\partial \sigma}\right)^2\!(\Delta \sigma)^2 \!+\! \left(\frac{\partial F_p}{\partial \gamma}\right)^2 \!(\Delta \gamma)^2\!+\!\left(\frac{\partial F_p}{\partial d}\right)^2\!(\Delta d)^2,
\end{equation}
where the partial derivatives, such as \(\frac{\partial F_p}{\partial \sigma}\), are computed numerically at each \(t_i\) using:

\begin{equation}
    \frac{\partial F_p}{\partial \sigma} = \frac{F_p(\hat{\sigma} + \delta \sigma, \hat{\gamma}, \hat{d}, t_i) - F_p(\hat{\sigma} - \delta \sigma, \hat{\gamma}, \hat{d}, t_i)}{2\delta \sigma},
\end{equation}
where \(\delta \sigma\) is a small perturbation. Similar procedures are applied for \(\frac{\partial F_p}{\partial \gamma}\) and \(\frac{\partial F_p}{\partial d}\).

For a time trace curve with \(N > 3\) points (\(t_1, t_2, \dots, t_N\)), the error propagation matrix equation can be written as:

\[
\begin{bmatrix}
\big(\!\Delta F_p(t_1)\!\big)^2 \\
\big(\!\Delta F_p(t_2)\!\big)^2 \\
\big(\!\Delta F_p(t_3)\!\big)^2 \\
\vdots \\
\big(\!\Delta F_p(t_N)\!\big)^2
\end{bmatrix}
\!=\!
\begin{bmatrix}
\partial_\sigma F_p(t_1) & \partial_\gamma F_p(t_1) & \partial_d F_p(t_1) \\
\partial_\sigma F_p(t_2) & \partial_\gamma F_p(t_2) & \partial_d F_p(t_2) \\
\partial_\sigma F_p(t_3) & \partial_\gamma F_p(t_3) & \partial_d F_p(t_3) \\
\vdots & \vdots & \vdots \\
\partial_\sigma F_p(t_N) & \partial_\gamma F_p(t_N) & \partial_d F_p(t_N)
\end{bmatrix}\!
\begin{bmatrix}
(\!\Delta \sigma)^2 \\
(\!\Delta \gamma)^2 \\
(\!\Delta d)^2
\end{bmatrix}.
\]
This matrix can be written compactly as : $\overrightarrow{\Delta F} = [\partial F_p] \overrightarrow{\Delta x}$, where the least-squares solution for \(\overrightarrow\Delta x\) is given by:

\begin{equation}
    \overrightarrow{\Delta x} = \left([\partial F_p]^T [\partial F_p]\right)^{-1} [\partial F_p]^T \overrightarrow{\Delta F},
\end{equation}
where \([\partial F_p]^T\) represents the transpose of the matrix. 

\begin{figure}[ht]
    \centering
    \includegraphics[width=0.48\textwidth]{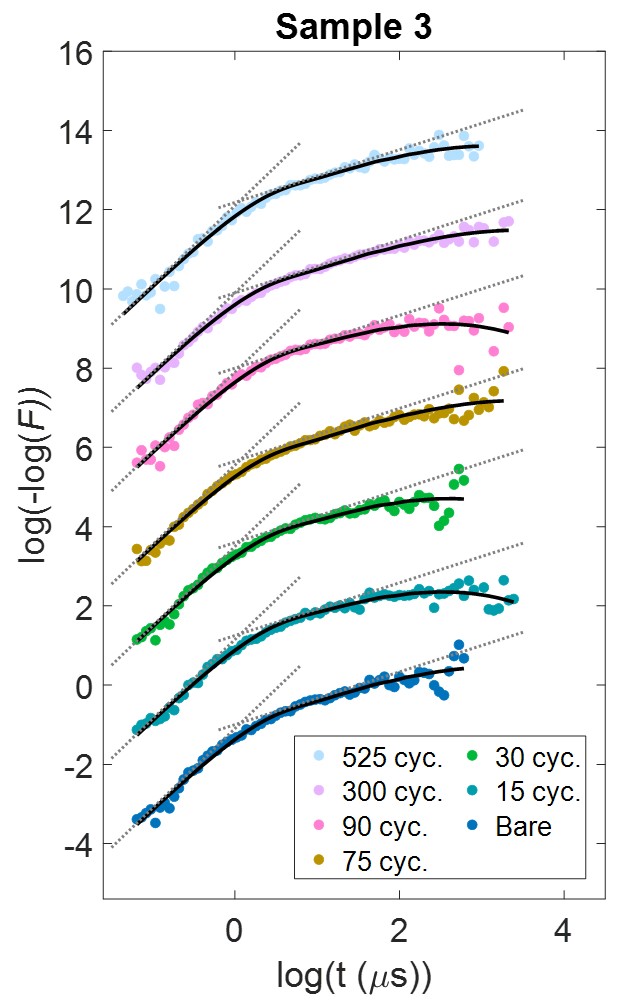}
    \caption{FID curves for Sample 3. Data fits well with 2D bath model (solid black line). Slope curves of $n = 2$ and $n = 2/3$ are included for reference (dotted lines).}
    \label{figSI:Samp13FID}
\end{figure}

Fig.~\ref{fig:FIDModel}(b,c) show the FID curve fitting for Samples 1 and 2. The corresponding data for Sample 3 is given in Fig.~\ref{figSI:Samp13FID}. All coating curves fit well with a 2D bath model due to the incomplete elimination of surface dark spins.

\section{NV Center Measurements}

\subsection{NV Photoluminescence}\label{NVPLappendix}
\begin{figure}[!ht]
    \centering
    \includegraphics[width=0.48\textwidth]{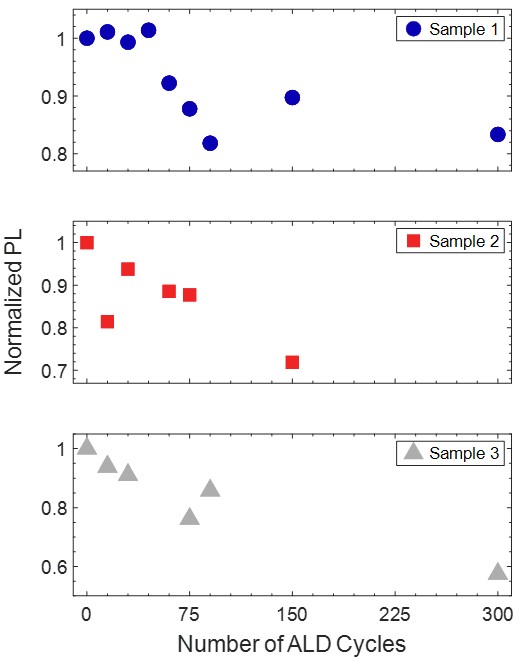}
    \caption{NV PL for the three ensemble NV diamond samples. Each value is normalized to the PL for the bare diamond case. All data for a given diamond is averaged an equal number of times to visualize the trend as a function of \ch{TiO2} thickness.}
    \label{figSI:NVPL}
\end{figure}

To assess how the \ch{TiO2} coatings affect the NV photoluminescence (PL), we analyze the raw DEER data comprising Fig.~\ref{fig:DEER}(c) for Samples 1-3. Since the raw data for each Hahn-echo background subsequence is relatively flat, we average over all frequencies ($x$-axis) to get a fluorescence level for each experimental condition. We plot the NV PL normalized to the bare diamond case in Fig.~\ref{figSI:NVPL}. For each diamond, all coating datasets are averaged the same number of times. All three NV ensemble samples show diminishing PL as the \ch{TiO2} layer gets thicker. 

\subsection{NV $T_1$} \label{NVT1Fit}
\begin{figure}[!ht]
    \centering
    \includegraphics[width=0.45\textwidth]{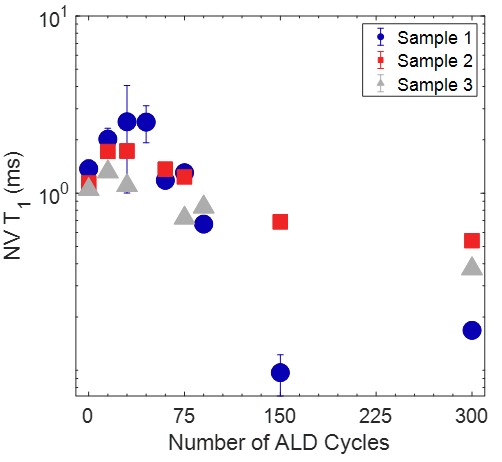}
    \caption{NV $T_1$ measurement. Island growth regime demonstrates an increase in $T_1$ relaxation time followed by an order of magnitude decrease for conformally grown films.}
    \label{figSI:NVT1}
\end{figure}

The NV $T_1$ as a function of \ch{TiO2} thickness was measured and plotted in Fig.~\ref{figSI:NVT1}. These values (as well as their stretching factors) are used in the fitting of the dark spin $T_1$ data (Fig.~\ref{fig:DarkT1}(b)). More details are provided in Appendix \ref{SI:DarkSpinT1}.

\subsection{NV $T_2$}\label{NVT2appendix}

The Hahn-echo decoherence of NV centers arises from interactions with both surface electron spins and nearby nuclear spins (e.g., \textsuperscript{13}C). To analyze and fit the NV Hahn-echo data, we begin by modeling the case of an NV center coupled to a single nuclear spin. In the rotating frame of the NV electron spin, the Hamiltonian is given by

\begin{equation}
\mathcal{H} = \omega_n I_z + S_z \left( A I_z + B I_x \right) = \left( A m_s - \omega_n \right) I_z + B m_s I_x,
\end{equation}

\begin{figure}[hb]
    \centering
    \includegraphics[width=0.45\textwidth]{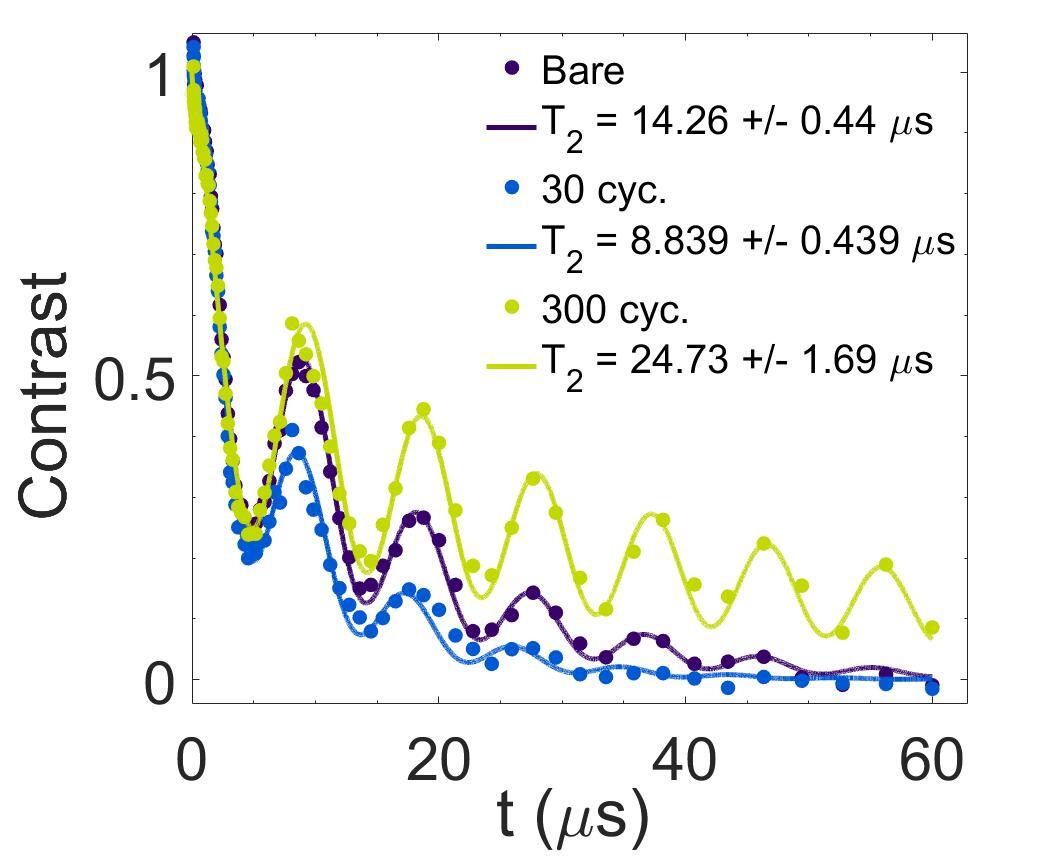}
    \caption{NV Hahn-Echo coherence for various coating thicknesses on Sample 1. Nuclear spin oscillations from lattice \textsuperscript{13}C can be seen for all coatings.}
    \label{fig:T2TimeTrace}
\end{figure}

\noindent where $m_s \in \{0, 1\}$ is determined by the spin state of the electron. Diagonalizing the Hamiltonian gives the resonance frequency $K = \sqrt{(A m_s - \omega_n)^2 + (B m_s)^2}$ such that $K_+ = \sqrt{(A - \omega_n)^2 + B^2} \quad \text{and} \quad K_- = \omega_n$.
The fitting function can thus be simplified to the following form~\cite{rowan1965electron}:
\begin{equation}
\begin{split}
&f_{\mathrm{Echo}}(\tau)\\
&=\!A e^{-(\frac{t}{\mathrm{T_2}})^n} \!\left[ \!1 \!+\!2 \Big(\!\frac{B}{K_+}\!\Big)^2 \!\sin^2\!\Big(\!\frac{\pi K_+ t}{2}\!+\!\phi_0 \!\Big) \!\sin^2\!\Big(\!\frac{\pi \omega_n t}{2}\!+\!\phi_1\!\Big) \right].
\end{split}
\end{equation}

In ensemble NV samples, the random spatial distribution of \textsuperscript{13}C spins leads to inhomogeneous averaging over the fast oscillation term involving $K_+$. The $2\left(\frac{B}{K_+}\right)^2\sin^2(\pi K_+/2w+\phi_0)$ term is averaged to a certain function $\alpha(t)$, accounting for the random distribution of nuclear spins. From the fittings, we realize that $\alpha(t)$ is approximately constant. Hence we have

\begin{equation}\label{eq:echofiiting}
    f(\tau)=A e^{-(\frac{t}{T_2})^n} \left[ 1 + \alpha \sin^2\left( \frac{\pi \omega_n t}{2} + \phi_1 \right) \right].
\end{equation}
The Sample 1 Hahn-echo time trace data is fitted using \ref{eq:echofiiting} and is plotted in Fig.~\ref{fig:T2TimeTrace} for a few coatings.

\begin{figure}[ht]
    \centering
    \includegraphics[width=0.45\textwidth]{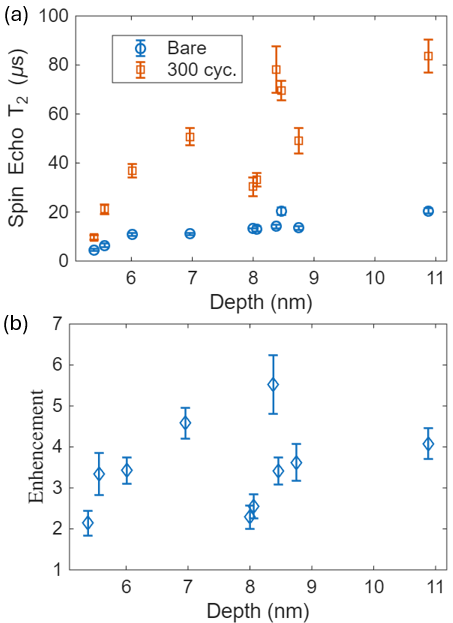}
   \caption{Sample 0 coherence measurements before (bare) and after 300 ALD cycles. (a) Hahn-echo $T_2$ versus depth, pre- and post-coating. (b) $T_2$ enhancement factor (coated/bare).}

    \label{fig:singleNVT2}
\end{figure}

For Sample 0, 10 individual NV centers are measured and shown in Fig.~\ref{fig:singleNVT2}. 
The depths for the bare case were measured by proton NMR~\cite{pham2016nmr}, as shown in Fig.~\ref{fig:singleNVT2}(a), and all ten NV centers exhibit an enhancement after coating as in Fig.~\ref{fig:singleNVT2}(b). To enable a direct comparison with the ensemble measurements (Samples~1–3) in the main text, we summed the Hahn-echo \(T_2\) time traces from the ten individual NV centers and fit the combined signal as an ensemble. 

\section{Dark Spin Correlation Measurements}\label{darkspinmeasurements}

\subsection{Dark Spin Rabi}
\begin{figure}[ht]
    \centering
    \includegraphics[width=0.48\textwidth]{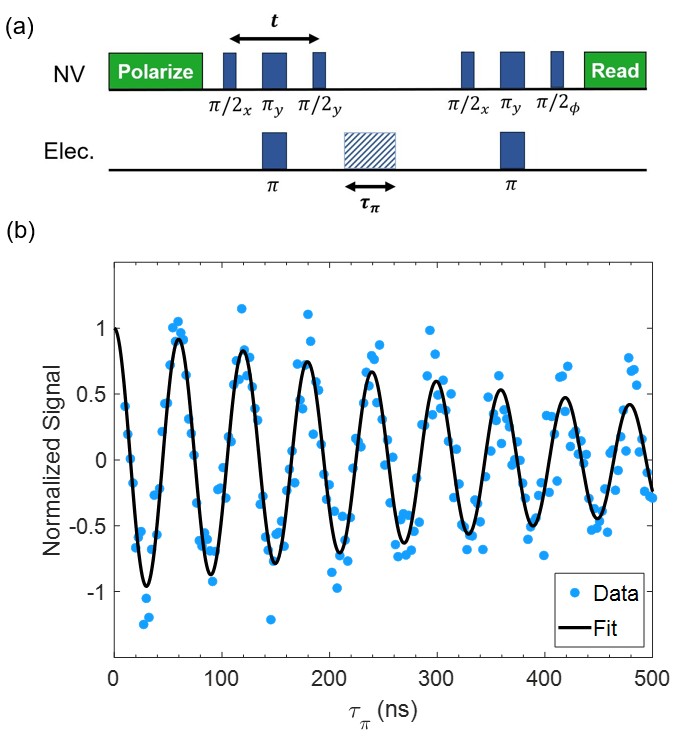}
    \caption{(a) DEER correlation-based Rabi sequence. Similar structure to DEER $T_1$ and $T_2$ sequence in Fig. 3 except the middle dark spin $\pi$ pulse duration is swept through a set of $\tau_\pi$ intervals. The duration of the first and last dark spin $\pi$ pulses in the two DEER subsequences is determined through a conventional DEER Rabi measurement. (b) DEER correlation-based Rabi data set for Sample 0.}
    \label{figSI:Corr-Rabi}
\end{figure}

The conventional way to determine dark spin Rabi is to vary the electron spin $\pi$ pulse duration in Fig.~\ref{fig:DEER}(a). However, if the interaction strength is large or phase accumulation time $t$ is large, it might induce additional faster oscillations which can cause the misidentification of the electron spin $\pi$ pulse duration ~\cite{zhang2021nanoscale}. Here we use a correlation type measurement in Fig.~\ref{figSI:Corr-Rabi} to avoid the problem~\cite{PhysRevLett.113.197601}.

The measurement result in Fig.~\ref{figSI:Corr-Rabi}(b) is fitted with decay sinusoidal function $e^{-(\tau_{\pi}/T_{\mathrm{Rabi}})^n}\cos(2\pi f \tau_{\pi})$, with $T_{\mathrm{Rabi}} = 545.3 \pm 129.8$ ns, $n =1.103\pm 0.8045$ , $f =16.7 \pm 0.04$ MHz.

\subsection{Nuclear Spin-Free Dark Spin $T_1$}
\label{SI:DarkSpinT1}
To directly measure the relaxation of surface electron spins, we modify the previous pulse sequence~\cite{PhysRevLett.113.197601}. The previous pulse sequence suffers from nuclear spin modulation due to the filter band width of the two dynamical decoupling subsequences.  In our case, the overall sequence is performed twice: once with a middle \(\pi\)-pulse and once without it (Fig.~\ref{fig:DarkT1}(a)). The difference between the two cancels the nuclear modulation. To see this, the free evolution is governed by the following Hamiltonian:

\begin{equation}
\mathcal{H}_r = \frac{\omega}{2} \sigma_z^{(n)} + \sigma_z^{(v)} \left(A_{\parallel}^{(n)} \sigma_z^{(n)} + A_{\perp}^{(n)} \sigma_x^{(n)} \right) + \frac{V_{dd}}{4} \sigma_z^{(v)} \sigma_z^{(e)}
\end{equation}

\noindent By considering the dephasing and relaxation channels of the system and environmental spins, the free evolution can be described by the master equation:

\begin{equation}
\begin{split}
\frac{d\rho}{dt} =& -i[\mathcal{H}_r, \rho] \\
&+ \frac{\gamma_{v_1}}{2} \left( \mathcal{D}[\sigma_+^{(v)}] + \mathcal{D}[\sigma_-^{(v)}] \right)
+ \frac{\gamma_{v_2}}{2} \mathcal{D}[\sigma_z^{(v)}]
\\&+ \frac{\gamma}{2} \left( \mathcal{D}[\sigma_+^{(e)}] + \mathcal{D}[\sigma_-^{(e)}] \right) 
+ \frac{\gamma_{e_2}}{2} \mathcal{D}[\sigma_z^{(e)}],
\end{split}
\end{equation}
which can be expressed as: $\frac{d\rho}{dt} = \mathcal{L}\rho$.

\noindent Here, \(v\), \(n\), and \(e\) label the system spin, nuclear spin, and environmental spin, respectively. The initial state is given by \(\rho(0) = \frac{1}{8} (\sigma_0^{(v)} + \sigma_z^{(v)}) \sigma_0^{(e)} \sigma_0^{(n)}\). Following the microwave pulse sequence shown in Fig. 3(a), free evolution occurs between any two pulses according to the master equation. At the end, measurements are made in the \(\sigma_z^{(v)}\) basis, with the signal \(F_\phi^A = \text{Tr}[\rho \sigma_z^{(v)}]\).

Given that the superoperator \(\mathcal{L}\) is a 64-by-64 matrix, diagonalizing \(e^{\mathcal{L}t}\) analytically is challenging and impractical. To simplify, we approximate the nuclear spin contribution as a classical AC magnetic field:

\begin{equation}
\mathcal{H}_r' = \frac{B}{2} \sin(\omega t + \chi) \sigma_z^{(v)} + \frac{V_{dd}}{4} \sigma_z^{(v)} \sigma_z^{(e)}.
\end{equation}

By solving the $T_1$ pulse sequence in Fig.~\ref{fig:DarkT1}(a) with and without the middle \(\pi\)-pulse on the electron spin, and then subtracting the two results, we obtain:
\begin{widetext}
\begin{equation}
F_y^\pi \!-\! F_y^0 \!=\; 
    e^{-(\gamma_{v_1} + \gamma)\tau} \!\bigg[ e^{-\left(\frac{\gamma_{v_1} + \gamma}{2} + \gamma_{v_2}\right)t}\!
    \frac{|V_{dd}|}{\sqrt{\!V_{dd}^2 \! - \!\gamma^2}\!} \sin\!\left(\!\frac{t}{2}\! \sqrt{\!V_{dd}^2 \!-\! \gamma^2\!} \right) \bigg]^2 
    \propto  \, e^{-(\gamma_{v_1} + \gamma)\tau},
\end{equation}
\end{widetext}
where the AC magnetic field modulation information $\omega$ is not contained in the result. 

\begin{figure}[ht]
    \centering
    \includegraphics[width=0.4\textwidth]{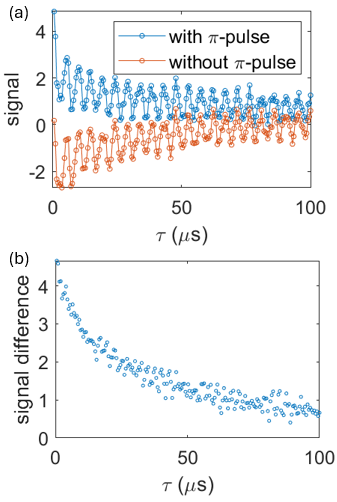}
    \caption{Correlation $T_1$ measurement on electron spins. (a) With $\pi$-pulse (blue), without $\pi$-pulse (red), both show nuclear modulation due to \textsuperscript{13}C. (b) The subtraction removes the nuclear modulation and what remains is a decay curve due to NV $T_1$ and electron spin $T_1$. }
    \label{figSI:ET1_experimental}
\end{figure}

We also perform the sequence altering the final \(\pi/2\)-pulse to be along the -\(y\)-axis, yielding: $F_{-y}^\pi - F_{-y}^0 \propto -e^{-(\gamma_{v_1} + \gamma) \tau}$. The measurement demonstration is shown in Fig.~\ref{figSI:ET1_experimental}, where (a) shows the nuclear spin modulation with and without $\pi$-pulse and (b) shows that the subtraction removes the modulation and the result is a decay curve due to NV $T_1$ and electron spin $T_1$. By subtracting the two signals, we calculate

\begin{equation}\label{eq:T1-sequence}
\text{S} = \left(F_y^\pi - F_y^0\right) - \left(F_{-y}^\pi - F_{-y}^0\right)\propto e^{-(\gamma_{v_1} + \gamma) \tau},
\end{equation}
which cancels systematic errors due to instrument control. 
Eq.~\ref{eq:T1-sequence} assumes the noise on the dark electron spins is Markovian. We fit the dark spin $T_1$ data in Fig.~\ref{fig:DarkT1}(b) with a more general form given by 

\begin{equation}\label{eq:darkspinT1Fit}
    f(\tau) = Ae^{-(\tau/T_{1,\mathrm{NV}})^{n_{\mathrm{NV}}}-(\tau/T_{1,e})^{n_{e}}} + c,
\end{equation}
where $n_{NV}$ and $n_{n_e}$ are stretch factors depending on the noise environment. To extract dark spin information, we use the $T_{1,NV}$ and $n_{NV}$ from Fig.~\ref{figSI:NVT1} and fit the electron $T_{1,e}$ and $n_{e}$ values for the various coatings using Eq.~\eqref{eq:darkspinT1Fit}. The results in Fig.~\ref{fig:DarkT1}(b) show the highlighted decrease in dark spin $T_1$.

\subsection{Dark Spin $T_2$}\label{appendix:darkspinT2}
\begin{figure}[ht]
    \centering
    \includegraphics[width=0.49\textwidth]{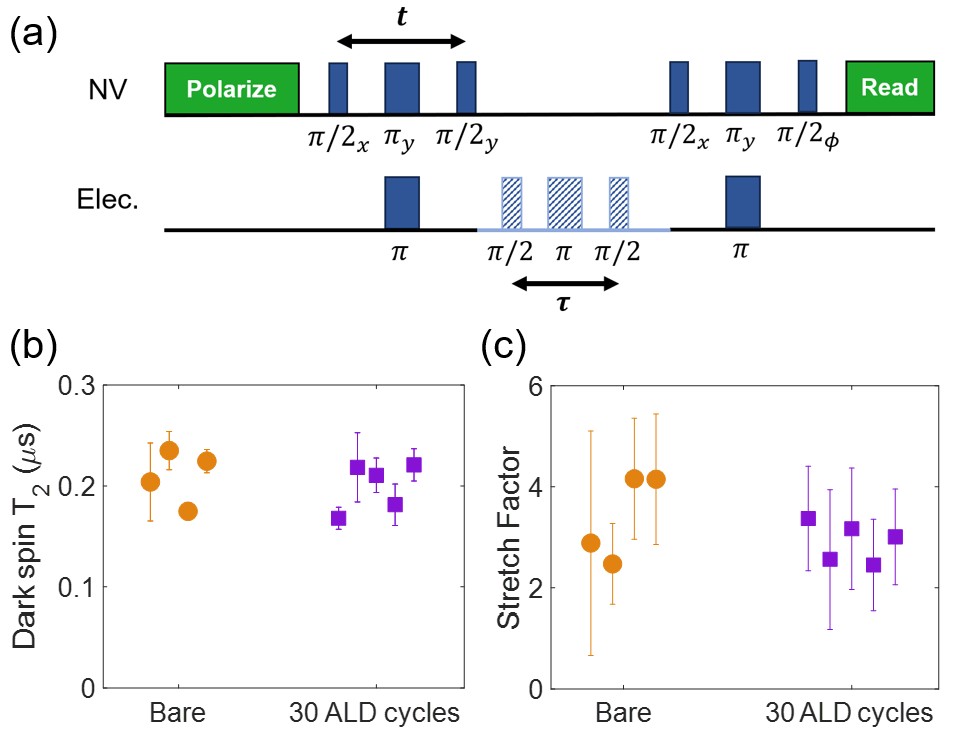}
    \caption{Dark spin $T_2$ measurement. (a) Correlation pulse sequence for probing dark spin $T_2$. (b) Dark spin $T_2$ measurement for Sample 0 before and after coating with 30 cycles of \ch{TiO_2}. (c) Corresponding stretching factors for DEER $T_2$ data in (b).}
    \label{figSI:DarkT2}
\end{figure}

For the dark spin $T_2$ measurement, the pulse sequence is shown in Fig.~\ref{figSI:DarkT2}(a). It is similar in structure to the sequence for dark spin $T_1$ (Fig.~\ref{fig:DarkT1}(a)), only the middle dark spin $\pi$ pulse is replaced with a Hahn-Echo sequence. The separation between the two probe segments is fixed, while $\tau$ is varied within the evolution segment.

Fig.~\ref{figSI:DarkT2}(b) shows dark spin $T_2$ times for four distinct NV centers in Sample 0 before and after coating with 30 cycles of \ch{TiO_2}. 
Interestingly, the ALD coating does not alter the dephasing rate of the dark spins suggesting that the dephasing mechanism is independent of \ch{TiO2}-thickness and not limited by defects in this \ch{TiO_2} capping layer.
The stretching factor for the bare case is 3.42 $\pm$ 1.47 and for 30 ALD cycles is $n_\mathrm{e} =2.91 \pm 1.11$ (Fig.~\ref{figSI:DarkT2}(c)). These values are not significantly different from each other and indicate the decoherence is mainly due to nuclear spins~\cite{yang2016quantum}, namely the proton spins in the objective immersion oil.  

\begin{figure}[ht]
    \centering
    \includegraphics[width=0.48\textwidth]{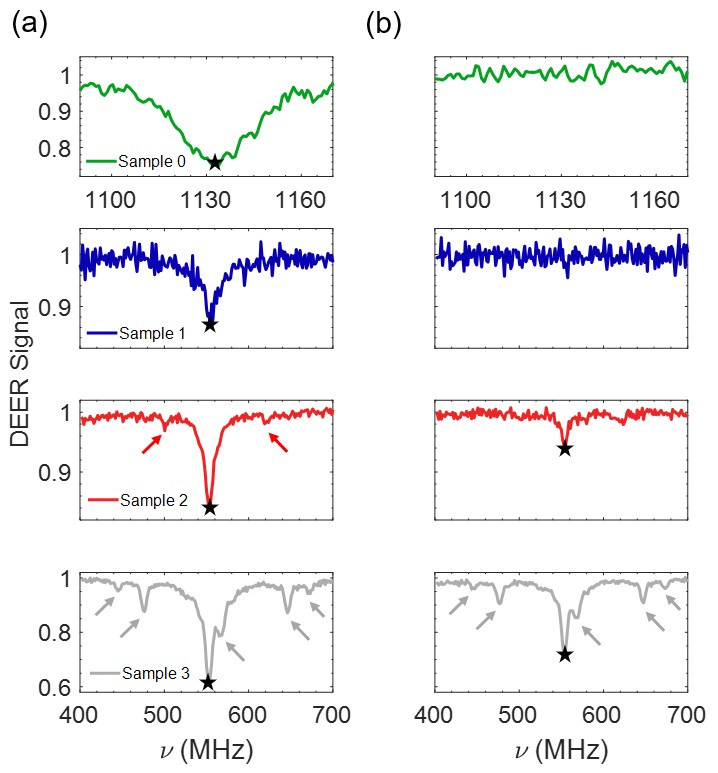}
    \caption{DEER profile comparison between bare (a) and 300 ALD cycles (b). The four samples show smaller reductions in the dark spin resonance (black star) as the nitrogen implantation dose in increased from Sample 0 (green) to Sample 3 (gray). The appearance of P1 resonances (arrows) becomes apparent at higher dosages in Sample 2 (red) and Sample 3 where the splittings correspond to \textsuperscript{15}N and \textsuperscript{14}N for the two samples respectively.}
    \label{figSI:DEERProfiles}
\end{figure}

\begin{figure*}
    \centering
    \includegraphics[width=\textwidth]{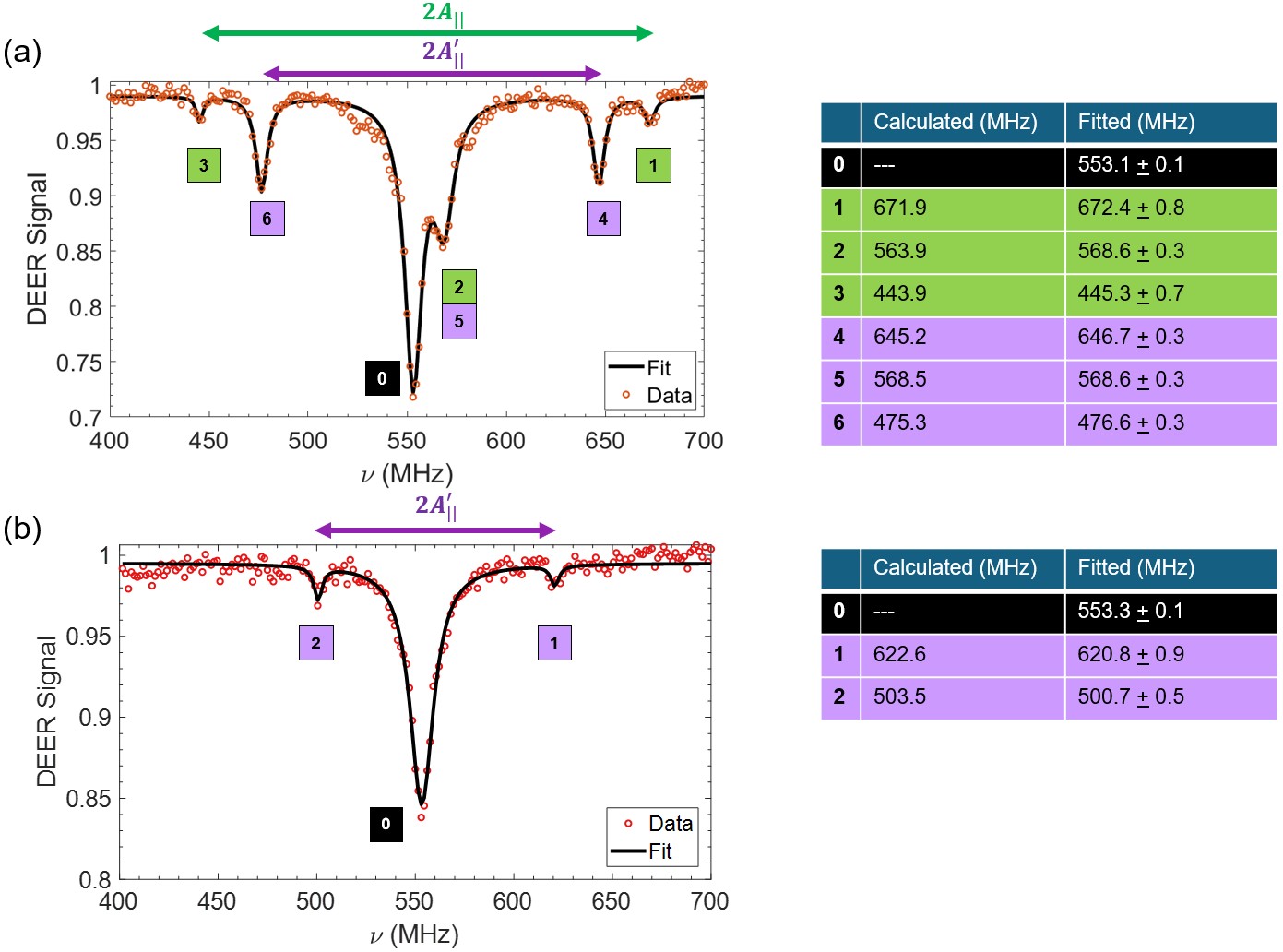}
    \caption{Identification of P1 peaks for high dosage samples. (a) DEER spectrum for Sample 3 coated with 525 ALD cycles, composed of \textsuperscript{14}N. Numbered peaks are fit with a six-term Lorentzian and show good agreement with the theoretical values found from equation \ref{eq:P1_14N}. Green (purple) peaks represent on-axis (off-axis) P1 center resonances. The discrepancy for peak 2 comes from our inability to spectrally resolve $\omega_{t_2}$ (on-axis) and $\omega'_{t_2}$ (off-axis), leading to the choice of a six-term fit instead of a seven-term fit. (b) DEER spectrum for Sample 2 with no coating, composed of \textsuperscript{15}N. Similar table shown on right comparing the fitted values of the three-term Lorentzian with the values computed in equation \ref{eq:P1_15N}.}
    \label{figSI:gfactor}
\end{figure*}

\section{P1 Center DEER Signals}\label{P1}
Since P1 centers consist of a single unpaired electron, they too can be probed by DEER techniques \cite{PhysRevB.94.155402}. 
We observe the appearance of P1 center DEER signals as the dose of nitrogen implantation is increased across the four samples. 
Fig.~\ref{figSI:DEERProfiles}(a) shows the bare diamond DEER profile for each sample. The surface spin resonance (star) is clearly identifiable and can be driven with MW to assess its contribution to the NV decoherence. 
Fig.~\ref{figSI:DEERProfiles}(b) shows the corresponding DEER spectra for each diamond coated with 300 ALD cycles and reveals the varying degrees of residual dark spin signals after coating. Sample 0 shows a complete reduction of the dark spin resonance from $23.99 \pm 0.43$\% to undetectable. Sample 1 similarly shows a complete reduction from $10.08 \pm 0.53$ \% to undetectable. The reduction for Sample 2 is from $14.92 \pm 0.26$\% to $5.26 \pm 0.32$\% and for Sample 3 is from $32.72 \pm 0.75$\% to $24.96 \pm 0.47$\%. 

The DEER spectra from Sample 2 (Fig.~\ref{figSI:DEERProfiles} red) and Sample 3 (Fig.~\ref{figSI:DEERProfiles} gray) show clear examples of capturing both dark spin defects and bulk P1 defects. 
The emergence of additional signals (identified by arrows) off resonance from the dark spin is indicative of the higher density of P1 centers in Samples 2 and 3. These signals correspond to single transitions where the P1 center electron is flipped while its nuclear spin remains constant. 
The frequencies of the peaks are governed by the hyperfine coupling terms $A_{||}$ and $A_{\perp}$ in the P1 center Hamiltonian which describe P1 centers with their crystallographic orientation aligned (on-axis) and misaligned (off-axis) with the external B field, respectively \cite{PhysRevB.94.155402}. 
The Hamiltonian for a \textsuperscript{14}N P1 center is given by 
\begin{equation}
    \mathcal{H}_{P1} = \omega_e S_z+ \omega_n I_z+A_\parallel S_zI_z +A_\perp(S_xI_x+S_yI_y) + QI_z^2
\end{equation}
where $\omega_e$ and $\omega_n$ are the Larmor precession frequencies of the P1 electron and nuclear spin respectively, $\mathbf{S} = (S_x, S_y, S_z)$ and $\mathbf{I} = (I_x, I_y, I_z)$ are the electron and nuclear spin operators, and $Q$ is the quadrupolar coupling constant for the nuclear spin \cite{PhysRevB.94.155402}. For \textsuperscript{15}N, there is no nuclear quadrupole moment and the last term is omitted. 
The on-axis P1 centers exhibit larger splittings as they experience the full magnitude of the external B field and are smaller in magnitude as they only account for approximately one fourth of all P1 centers in the diamond. 
The expressions for the on-axis P1 center resonances are obtained by diagonalizing $H_{P1}$. For \textsuperscript{14}N, they are \cite{PhysRevB.94.155402}, 


\begin{align}\label{eq:P1_14N}
\omega_{t_1}&=\omega_t \left( \left| + \frac{1}{2}, +1 \right\rangle \to \left| -\frac{1}{2}, +1 \right\rangle \right) = \omega_e + A_\parallel + \frac{A_\perp^2}{2 \omega_e} \nonumber \\
\omega_{t_2}&=\omega_t \left( \left| + \frac{1}{2}, 0 \right\rangle \to \left| -\frac{1}{2}, 0 \right\rangle \right) \quad\;\;= \omega_e + \frac{A_\perp^2}{\omega_e} \nonumber \\
\omega_{t_3}&=\omega_t \left( \left| + \frac{1}{2}, -1 \right\rangle \to \left| -\frac{1}{2}, -1 \right\rangle \right) = \omega_e - A_\parallel + \frac{A_\perp^2}{2 \omega_e}
\end{align}

\noindent while for \textsuperscript{15}N 

\begin{align}\label{eq:P1_15N}
\omega_{t_1}\!=\!\omega_t \Big( \big| \!+ \frac{1}{2}, \!+\frac{1}{2} \big\rangle \!\to\! \big|\! -\frac{1}{2}, \!+\frac{1}{2} \big\rangle \Big) \!&=\!
\omega_e\!+\!\frac{A_\parallel}{2}\!+\! \frac{A_\perp^2}{4 (\omega_e\!-\!\omega_n)} \nonumber \\
\omega_{t_2}\!=\!\omega_t \Big( \big|\!+\frac{1}{2}, \!-\frac{1}{2} \big\rangle \!\to\! \big| \!-\frac{1}{2}, \!-\frac{1}{2} \big\rangle \Big) \!&= \!
\omega_e \!-\! \frac{A_\parallel}{2} \!+\! \frac{A_\perp^2}{4(\omega_e\!-\!\omega_n)}
\end{align}

\noindent To compute the off-axis resonances, the associated coupling constants $A'_{\parallel}$ and $A'_{\perp}$ are used and are related to the on-axis constants by
$ A'_{\parallel} = \frac{1}{9} (A_{\parallel} + 8 A_{\perp})$ and $
A'_{\perp} = \frac{1}{9} (4 A_{\parallel} + 5 A_{\perp})$ \cite{PhysRevB.94.155402}.

Fig.~\ref{figSI:gfactor} shows a representative DEER spectrum for Sample 3 and Sample 2 in more detail as well as the theoretical and fitted values for the resonances. The theoretical values are found using the above equations and the fitted values come from six-peak Lorentzian fits for Sample 3 and three-peak Lorentzian fits for Sample 2. In Fig.~\ref{figSI:gfactor}(a), Sample 3 has on-axis (green - peaks 1, 2, 3) and off-axis (purple - peaks 4, 5, 6) P1 center resonances visible. The on-axis splitting is found to be $227.1 \pm 1.1$ MHz from the fitting and the off-axis splitting is $170.1 \pm 0.4$ MHz. These splittings agree with values for \textsuperscript{14}N.
In contrast, Sample 2 exhibits two peaks (labeled 1 and 2) split by $120.1 \pm $ MHz (from curve fit), which is in good agreement with the theoretical prediction for off-axis \textsuperscript{15}N P1 centers (Fig.~\ref{figSI:gfactor}(b)). The on-axis P1 centers are not detectable in this case. 

Furthermore, the splitting near 560 MHz for Sample 3 is characteristic of \textsuperscript{14}N implantation, which produces three distinct resonances each for on and off-axis P1 centers from the $I = 1$ nuclear spin. Here, the surface spin (precessing at the electron Larmor frequency $\omega_e$) is spectrally resolved from the $m_I = 0$ hyperfine transition of the \textsuperscript{14}N P1 center (precessing at $\omega_e + \frac{A_{\perp}^2}{\omega_e}$) \cite{PhysRevB.94.155402}. Conversely, the lack of splitting for Sample 2 is indicative of the $I = 1/2$ nuclear spin in \textsuperscript{15}N which splits into two hyperfine resonances. 

As mentioned in the main text, the NV-P1 coupling does not influence the dark spin experiments performed due to their distinct spectral signature. For Sample 2, the \textsuperscript{15}N P1 centers' signals are separated from the dark spin resonance by more than 50 MHz, while for Sample 3, the middle \textsuperscript{14}N P1 resonance is separated by 10 MHz. Given our dark spin Rabi drive at 4.17 MHz, maximally we could only be driving 15\% of the P1 population. This is an overestimate, however, as our Rabi frequency is not optimized for P1 driving. 

\section{g-factor estimation }\label{g-factor}
We use the bare diamond DEER spectrum from Sample 3 to compute the g-factor of the dark spins under investigation. The spectrum in Fig.~\ref{figSI:gfactor}(a) is fit with a six-term Lorentzian function, one peak for the dark spin signal and five peaks for the P1 signals. Note that peaks 2 and 5 in the spectrum correspond to the on-axis and off-axis $\omega_{t2}$ resonances and overlap due to microwave power broadening. These two peaks are fit as one and explain why the on-axis peak 2 in the table in Fig.~\ref{figSI:gfactor}(a) shows some discrepancy from the fitted value while peak 5 fits well.
We start by using our fitted values for $\omega_{t_1}$, $\omega_{t_2}$, and $\omega_{t_3}$ to solve

\begin{equation}
     \omega_{t_1}+\omega_{t_3}-\omega_{t_2} = \omega_{e} 
\end{equation}
to isolate $\omega_e$. We compute $\omega_e$ for both the on-axis (peaks 1,2,3) and off-axis P1 resonances (peaks 4,5,6) and obtain values of 549.1 MHz and 554.7 MHz respectively. We average together the results to get a final $\omega_{e,avg} = 551.9$ MHz. To solve for $g$ of the dark spins, we use $\omega_{e,avg}$ to find the effective magnetic field in the experiment $B_{eff}$ 

\begin{equation}
    B_{eff} = \frac{\omega_{e,avg}}{\gamma_e}
\end{equation}
where $\gamma_e = 2.8024$ MHz/G and obtain $B_{eff} \approx 196.9$ G. Lastly, we solve for the g-factor of the spin resonance using 

\begin{equation}
    g = \frac{2\pi\hbar\omega_{e,fit}}{\mu_B B_{eff}}.
\end{equation}
Here, $\omega_{e,fit}$ is the fitted value for the dark spin peak (Fig.~\ref{figSI:gfactor}(a), labeled 0) and $\mu_B$ is the Bohr magneton. From this, we find $g = 2.0067(21)$.

The process is similar for using the data from Sample 2 to compute the g-factor. We start by using our fitted values for $\omega_{t_1}$ and $\omega_{t_2}$ (Fig.~\ref{figSI:gfactor}(b)) to solve 

\begin{equation}
    \omega_{t_1} + \omega_{t_2} = 2\omega_e + \frac{A_\perp^2}{2\omega_e}.
\end{equation}
where we assume $\omega_e \gg \omega_n$. By summing $\omega_{t1}$ and $\omega_{t2}$, we can solve for the dark spin Larmor frequency $\omega_e$ and extract the effective magnetic field $B_{eff}$ experienced by the dark spins in the experiment. Rearranging, we find the quadratic expression for $\omega_e$
\begin{equation}
    \omega_e^2 - \frac{\omega_{t_1} + \omega_{t_2}}{2}\omega_e + \frac{A_\perp^2}{4} = 0
\end{equation}
which can be solved by the quadratic formula
\begin{equation}
    \omega_e = \frac{\frac{\omega_{t_1} + \omega_{t_2}}{2} \pm \sqrt{\frac{(\omega_{t_1} + \omega_{t_2})^2}{4} - A_\perp^2}}{2}.
\end{equation}
We find $\omega_e = 549.0944$ MHz and $\omega_e = 11.6556$ MHz, where the first root fits in the frequency range of interest in our measurements. Following the same procedure as for Sample 3, we end up with a result $g = 1.9966(19)$.
%


\section{EPR Sensitivity}\label{EPR Sensitivity Cal}
We estimate the impact of our surface passivation technique on nanoscale NV-EPR sensitivity. Our surface modification strategy has two effects. On the one hand, the \ch{TiO_2} layer reduces the dark spin density and improves the NV coherence, which enhances the quantum sensor's sensitivity. On the other hand, the \ch{TiO_2} layer increases the distance  between the NV center and the target, which in turn reduces dipolar coupling strength. 

\begin{figure}[!hb]
    \centering
    \includegraphics[width=0.46\textwidth]{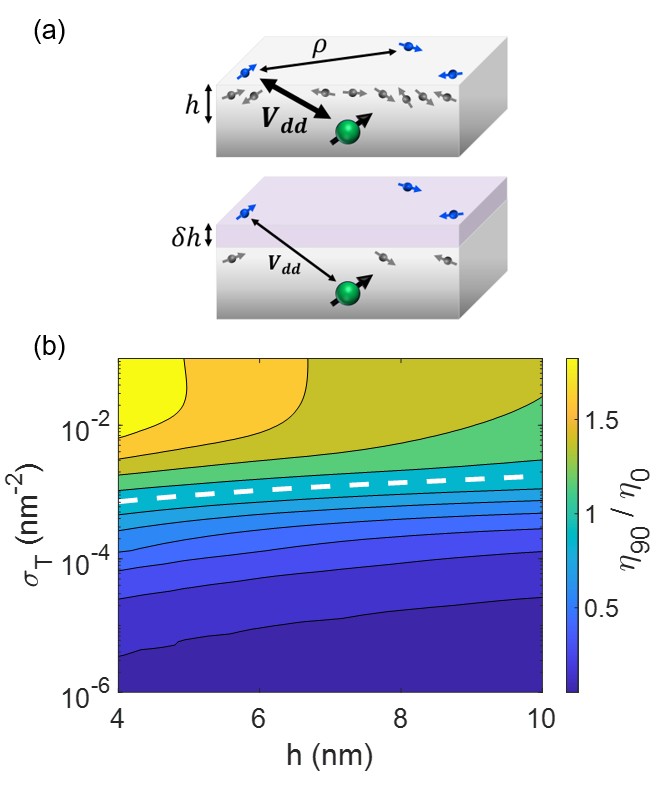}
    \caption{ALD coating effects on EPR sensitivity. (a) Schematic of NV EPR without (top) and with (bottom) \ch{TiO2} coating. Target spins (blue) are sensed by the NV center (green) through dipolar coupling ($V_{dd}$). Although the dipolar coupling is weaker when an additional separation $\delta h$ is introduced from the \ch{TiO2} film, the dark spin (gray) density is reduced, leading to certain conditions for increased EPR sensitivity. (b) Simulation depicting the impact of \ch{TiO2} coating on NV EPR sensitivity. The contours represent the ratio in sensitivities of 90 and 0 ALD cycles for varying NV depths ($h$) and target spin densities ($\sigma_T$). Blue regions on the contour plot illustrate experimental conditions for increased sensitivity with the coated diamond. The dashed white line represents a ratio of 1.}
    \label{figSI:NVEPR}
\end{figure}

If the target molecule grafting density on the diamond sensor is low (Fig. \ref{figSI:NVEPR}(a)), the addition of a thin \ch{TiO_2} layer of 90 cycles results only in a small change of the NV-target separation and, therefore, only marginally impacts $V_{dd}$. Specifically, if we assume that the average spacing between target molecules ($\rho$) is large compared to the NV-surface distance ($h$), an added \ch{TiO_2} layer with thickness $\delta h$ results in only a modest change in dipolar interaction strength $\frac{\delta V_{dd}}{V_{dd}} \approx - \frac{3 h}{\rho^2} \delta h$. At the same time, using Sample 1 as an example, the coating leads to an over five times decrease in dark spin density $\sigma$ (see Fig.~\ref{fig:FIDModel}(d)) which largely eliminates signal from non-target spins. To assess this trade-off, Fig.~\ref{figSI:NVEPR}(b) shows a density plot indicating the sensitivity ($\eta$) increase of a diamond surface coated with 90 ALD cycles ($\sim4 \mathrm{nm}$ \ch{TiO2}) compared to that of an uncoated surface as a function of NV depth and target spin grafting density ($\sigma_T$). We find that for typical molecular biological grafting densities of less than $1,000$~$\mu$m$^{-2}$ \cite{PNAS.119.8}, a 90 cycle \ch{TiO_2} layer results in an overall EPR sensitivity enhancement. This suggests that the benefit of reducing the dark spin density $\sigma$ outweighs the larger separation between the sensor and target spin, paving the way for superior NV EPR performance in biological experiments.

To evaluate the sensitivity to target electron spins (with $g=2$) at density $\sigma_T$, we define the minimal detectable density as

\begin{equation}\label{minideltasigma}
\delta \sigma_T = \frac{1}{M} \frac{\Delta S_z}{\frac{d\langle S_z \rangle}{d\sigma_T}}.    
\end{equation}
Here, \(\Delta S_z\) and \(\langle S_z \rangle\) represent the standard deviation and expectation value, respectively, of the total signal (comprising contributions from both the target electron spins and the unwanted bath electron spins) detected by the NV center.
The standard deviation \(\Delta S_z\) is given by:

\begin{equation}
\Delta S_z = \sqrt{\text{Tr}(S_z^2 \rho) - \text{Tr}(S_z \rho)^2} = \sqrt{1 - \langle S_z \rangle^2}.
\end{equation}
The expectation value \(\langle S_z \rangle\) can be expressed as the product of noise contributions from the target and bath spins:

\begin{equation}
\langle S_z \rangle = e^{\sigma_T W(\gamma_T,d_T, \tau)} e^{\sigma_B W(\gamma_B,d_B, \tau)}.
\end{equation}
The number of repeated measurements is \(M = \frac{t_{\text{total}}}{\tau}\), where \(t_{\text{total}}\) is the total measurement time, and \(\tau\) is the duration of each measurement. By considering the total duration of the experiment, the sensitivity is then defined as:

\begin{equation}
\eta = \delta \sigma_T \sqrt{t_{\text{total}}} = \frac{\Delta S_z \sqrt{\tau}}{\frac{d\langle S_z \rangle}{d\sigma_T}}
\end{equation}
For each data point, we search for the optimal \(\tau\) that minimizes \(\eta\). 

For a coating with 90 ALD cycles (i.e., thickness roughly 4 nm), the dark spin density begins to saturate. Thus, simulations were performed with a 4 nm coating (with $\gamma_B = 0.097$ MHz and $\sigma_B = 278.5$ $\mu\text{m}^{-2}$) and compared to the bare diamond case (Sample 1 from Fig.~\ref{fig:FIDModel}(d), \(\gamma_B = 0\) and \(\sigma_B = 1461 \, \mu\text{m}^{-2}\)). In the bare case, the target spin distance \(d_T\) is equal to the NV center depth. By contrast, for the 90 ALD cycles coating, the target spin is located 4 nm farther from the NV center, resulting in \(d_T = \text{NV depth} + 4 \, \text{nm}\). For both cases, the relaxation rate of the target spins is set to \(\gamma_T = 0\). 

\begin{figure}[ht]
    \centering
    \includegraphics[width=0.38\textwidth]{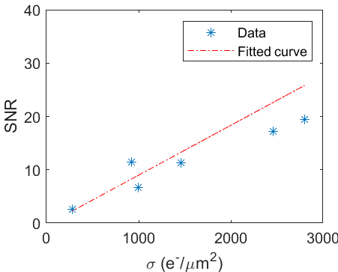}
    \caption{Signal-to-Noise Ratio (SNR) vs. electron spin density for Sample 2 FID data in Fig.~\ref{fig:FIDModel}(c). The data points (blue dots) are obtained by dividing the density by the standard deviation. The red line represents a linear fit given by $SNR = 0.0093\sigma - 0.3477$.}
    \label{SNRsigma}
\end{figure}

To estimate the experimental duration required to achieve a signal-to-noise ratio (SNR) of 1, we use real experimental parameters. We define the signal as the electron spin density and the noise as the standard deviation. For Sample 2 in Fig.~\ref{fig:FIDModel}(c), which presents a density plot, we compute the SNR for each data point. By plotting the SNR against spin density, we observe that the relationship is approximately linear, with a slope of \( a = 0.0093 \) and an intercept of \( b = -0.3477 \), as shown in Fig.~\ref{SNRsigma}. This establishes the relation between SNR and electron spin density. These results are obtained with an experimental duration of roughly 5 hours and an NV center depth of approximately 4 nm, as depicted in Fig.~\ref{fig:FIDModel}(d). If the density of the target spins is \( 500 \)~\(\mu\)m\(^{-2}\), we can compute the corresponding SNR and noise using the linear fitting. In the regime where interactions are weak, following Eq.~\ref{minideltasigma}, the noise contribution from target spins scales as \( \delta\sigma_T \propto d^2 \), where $d$ is the distance between NV and spin bath. The total noise is a combination of interface noise and target spin noise, added in quadrature. 

Thus, the SNR with coating is given by:

\begin{equation}
    SNR = \frac{\sigma_T}{\sqrt{\delta\sigma_{B}^2\frac{t_I}{t}+\left (\frac{\sigma_T}{a\sigma_T+b}\right )^2\frac{t_I}{t}\left (\frac{d_{nv}+h}{d_{nv}}\right )^4}},
\end{equation}
where \( h = 4 \) nm is the coating thickness, \( t_I = 5 \) hours is the experimental duration required to generate each data point in Fig.~\ref{SNRsigma}, and \( d_{nv} = 4 \) nm is the NV center depth. The background noise at a 4 nm coating is given by $\delta\sigma_B = 107.5~\mathrm{\mu m}^{-2}$. 

By setting \( SNR = 1 \), one can solve for \( t \). For Sample 1, \( t = 2.3 \) hours, whereas for Sample 2, which contains 10 times higher NV concentration, \( t = 15 \) minutes.

\section{Average minimum distance between NV and P1 center} \label{NV-P1 Dis Theory}

In this section, we estimate the average minimum NV--P1 center distance for each sample studied in the main text. We compute distances of 
$91.37 \pm 47.60~\mathrm{nm}$ (Sample~0), 
$11.93 \pm 5.59~\mathrm{nm}$ (Sample~1), 
$4.71 \pm 1.90~\mathrm{nm}$ (Sample~2), and 
$2.88 \pm 1.20~\mathrm{nm}$ (Sample~3). These proximities correspond to dipolar interaction strengths of roughly 
$60~\mathrm{Hz}$ (phase accumulation $\approx 0$), 
$27.35~\mathrm{kHz}$ (phase accumulation $\approx 0.04$), 
$0.44~\mathrm{MHz}$ (phase accumulation $\approx 0.7$), and 
$1.94~\mathrm{MHz}$ (phase accumulation $\approx 3.2$), respectively. 
Here, ``phase accumulation'' refers to the interaction-induced phase shift of the NV spin 
acquired during the DEER evolution time, expressed in radians. 
The large phase accumulation in Samples~2 and~3 is consistent with the clear observation 
of P1 resonance features in the DEER spectra (see Appendix~\ref{P1}). 

\begin{figure}[ht]
    \centering
    \includegraphics[width=0.35\textwidth]{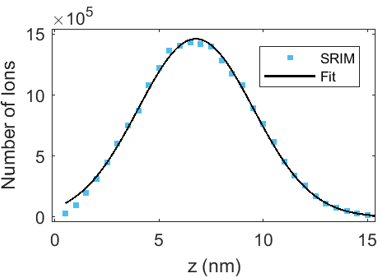}
    \caption{SRIM simulation showing number of implanted \textsuperscript{15}N ions as a function of depth in diamond from the surface. The data is fit to a Gaussian profile for the calculation.}
    \label{SRIM}
\end{figure}

The derivation of the distance between the $n^{th}$ nearest P1 center and an NV center starts with a Gaussian distribution on the $z$-axis and a uniform distribution on the $x$ and $y$ axes. The final result works for an arbitrary distribution and dimensions. Using a SRIM simulation assuming \textsuperscript{15}N implanted at 4 KeV and 5$^{\circ}$ tilt, the nitrogen density in diamond due to ion implantation is approximated from a Gaussian distribution on the z-axis (Fig. \ref{SRIM}) and uniformly distributed on x-y plane. The fit in the figure is Gaussian distribution with mean equal to $6.76 \pm 0.029$nm and standard deviation equal to $2.77\pm0.029$nm.

Next, we define the probability density function of nitrogen to be 

\begin{equation}
    p(x,y,z) =  \frac{1}{L^2} \frac{1}{\sqrt{2\pi\sigma^2}} e^{-\frac{(z-\mu)^2}{2\sigma^2}}.
\end{equation}

The depth of ion implantation is on the order of a few nanometers, which is much smaller than the thickness of the diamond sample, measured on the millimeter scale. Therefore, \( z \in \{0, +\infty\} \), while \( x \in \{0, L\} \) and \( y \in \{0, L\} \), where \( L \) represents the size of the diamond. Given that the mean depth is much larger than its standard deviation (e.g., \( \mu > 2\sigma \)), one can approximate the range of \( z \) to \( z \in \{-\infty, +\infty\} \).

If the NV center is at $(x',y',z')$, then the probability density function with respect to the NV is shifted as  

\begin{equation}
    p(x,y,z|x',y',z') =  \frac{1}{L^2} \frac{1}{\sqrt{2\pi\sigma^2}} e^{-\frac{(z-\mu+z')^2}{2\sigma^2}}
\end{equation}

\noindent we define $q$ to be the 2 dimensional ion implantation dosage density, which makes the total number of ions being implanted $N=qL^2$. By converting into spherical coordinates for a distance calculation, we have

\begin{equation}
    p(r,\theta,\phi|x',y',z') =  \frac{q}{N} \frac{1}{\sqrt{2\pi\sigma^2}} e^{-\frac{(r\cos{\theta}-\mu+z')^2}{2\sigma^2}}r^2\sin{\theta}
\end{equation}
with integration element $dM=dr d\theta d\phi$. Here, the sphere is centered at $(x',y',z')$.

The probability that the nearest P1 center is not within a sphere of radius $r$ centered at $(x',y',z')$ is

\begin{equation}
\begin{split}
    & P(\text{no spin is within r})\\
    =&\left[1-\int_0^r\int_0^{2\pi}\int_0^{\pi}p(r,\theta,\phi|x',y',z')d\theta d\phi dr\right]^N\\
    =&\left[1-\frac{q}{N}\int_0^r\int_0^{2\pi}\int_0^{\pi}\frac{1}{\sqrt{2\pi\sigma^2}} e^{-\frac{(r\cos{\theta}-\mu+z')^2}{2\sigma^2}}r^2\sin{\theta}dM\right]^N\\
    \approx&exp \left( -q \int_0^r\int_0^{2\pi}\int_0^{\pi}\frac{1}{\sqrt{2\pi\sigma^2}} e^{-\frac{(r\cos{\theta}-\mu+z')^2}{2\sigma^2}}r'^2\sin{\theta}dM\right)\\
    =&exp\left(-q\int_0^rf(r'|x',y',z')dr\right),
\end{split}
\end{equation}

\noindent In the third line, we assume that the total number of ions $N$ is large, so it can be converted into an exponential function. In the fourth line, we define 

\begin{equation} \label{f(r|xyz)}
    \begin{split}      f(r|x'\!,y'\!,z')\!=&\!\int_0^{2\pi}\!\int_0^{\pi}\!\frac{1}{\sqrt{2\pi\sigma^2}} e^{-\frac{(r\cos{\theta})\!-\!\mu\!+\!z')^2}{2\sigma^2}}r^2\!\sin{\theta}d\theta d\phi\\
    =&\pi r \Bigg(\! \mathrm{Erf}\bigg(\frac{r\!+\!\mu\!-\! z'}{\sqrt{2} \sigma}\bigg) + \mathrm{Erf}\bigg(\frac{r\! -\!\mu \!+\!z'}{\sqrt{2} \sigma}\bigg)\! \Bigg),
    \end{split}
\end{equation}
and the integration over $r$ is given by
\begin{widetext}
\begin{equation}\label{intf(r|xyz)}
\begin{split}
&\int_0^rf(r'|x'\!,y'\!,z'\!)dr'=\\
&\sqrt{\frac{\pi}{2}}\! 
\bigg ( e^{-\frac{(r\!+\!\mu\!-\!z')^2}{2 \sigma^2}}\!(r\!-\!\mu\!+\!z')\!
+\!e^{-\frac{(r\!-\!\mu\!+\!z')^2}{2 \sigma^2}} (r\!+\!\mu\!-\!z')\! \bigg)\!\sigma\!
+\!\frac{\pi}{2} \left(r^2\!-\!(\mu - z')^2\!-\!\sigma^2\right)\!
\Bigg( \!\mathrm{Erf}\!\bigg(\frac{r\!+\!\mu\!-\!z'}{\sqrt{2} \sigma}\bigg)\!
+\!\mathrm{Erf}\!\bigg(\!\frac{r\!-\!\mu\!+\!z'}{\sqrt{2} \sigma}\bigg)\!\Bigg).
\end{split}
\end{equation}
\end{widetext}

By realizing that the following equation holds:

\begin{equation}
\begin{split}
    P(\text{no spin is within r}) =& P(\text{first spin seen is $>$ r})\\
    =& 1-P(\text{first spin seen is $\leq$ r}),
\end{split}
\end{equation}
one then finds the cumulative probability of seeing the first spin, $ P(\text{first spin seen is $\leq$ r}) = 1-P(\text{no spin is within r})$. Hence, the probability density of the nearest spin can be computed:

\begin{equation}
\begin{split}
W(r|x',y',z') =& \frac{\partial P(\text{first spin seen is $\leq$ r})}{\partial r}\\
=&-\frac{\partial P(\text{no spin is within r})}{\partial r}
\end{split}
\end{equation}






\noindent which in turn means that the probability density is given by

\begin{equation}
    W(r|x',y',z')\!=q e^{\!-q\!\int_0^rf(r'|x',y',z')dr'}f(r|x',y',z').
\end{equation}
Here, the normalization condition $\int_0^\infty W dr =1$ is automatically satisfied.

By realizing the above formula is the probability density function for an inhomogeneous Poisson process distribution with first hitting time \cite{Siegrist_Poisson}, one can immediately readout the $n^{th}$ nearest spin distribution.

\begin{equation}
\begin{split}
    W_n(r|x',y',z') =&\frac{(q\int_0^rf(r'|x',y',z')dr')^{n-1}}{(n-1)!}\\ &\times qe^{-q\int_0^rf(r'|x',y',z')dr'}f(r|x',y',z')
\end{split}
\end{equation}

Our derivation is for a 3-dimensional and non-uniform case, and the result works for arbitrary dimensionality and distribution cases. For a 2-dimensional uniform case, by realizing that $f(r)=2\pi r$ and $q$ is the surface spin density, then $W_n=\frac{(\pi qr^2)^{n-1}}{(n-1)!}2\pi q r e^{-\pi q r^2}$ which recovers the previous result with the $n =1$ case ~\cite{PRXQuantum.3.040328}. For the 3-dimensional uniform case with $f(r) =4\pi r^2$ and $q$ the volume density, then $W_n = \frac{4\pi q r^2}{(n-1)!}\left( \frac{4\pi qr^3}{3}\right)^{n-1} e^{-\frac{4\pi qr^3}{3}}$ also recovers the previous result~\cite{hall2014analytic}.

The average of $n^{th}$ nearest distance at $(x',y',z')$ is then given by 

\begin{equation}
    l_n(x',y',z') = \int_0^\infty rW_n(r|,x',y',z')dr.
\end{equation}
For the 2D uniform case, $l_n=\frac{\Gamma(n+\frac{1}{2})}{(n-1)!\sqrt{\pi}\sqrt{q}}$, and for the 3D uniform case $l_n = \left(\frac{3}{4\pi}\right)^{\frac{1}{3}}\frac{\Gamma(n+\frac{1}{3})}{(n-1)!q^{1/3}} $.

The variance of the nearest distance can be computed as $V_n(x',y',z')=\int_0^\infty r^2W_n(r|,x',y',z')dr-(\int_0^\infty rW_n(r|,x',y',z')dr)^2$.

In our case where nitrogen is uniformly distributed on \( x \)-\( y \) and follows a Gaussian distribution along \( z \), the explicit forms of $f(r|x',y',z')$ and $\int_0^r f(r'|x',y',z')dr'$ are computed above (Eqs.~\ref{f(r|xyz)} and \ref{intf(r|xyz)}). Hence, the probability density function $W_n(r|x',y',z')$ is determined.

\begin{figure}[ht]
    \centering
    \includegraphics[width=0.48\textwidth]{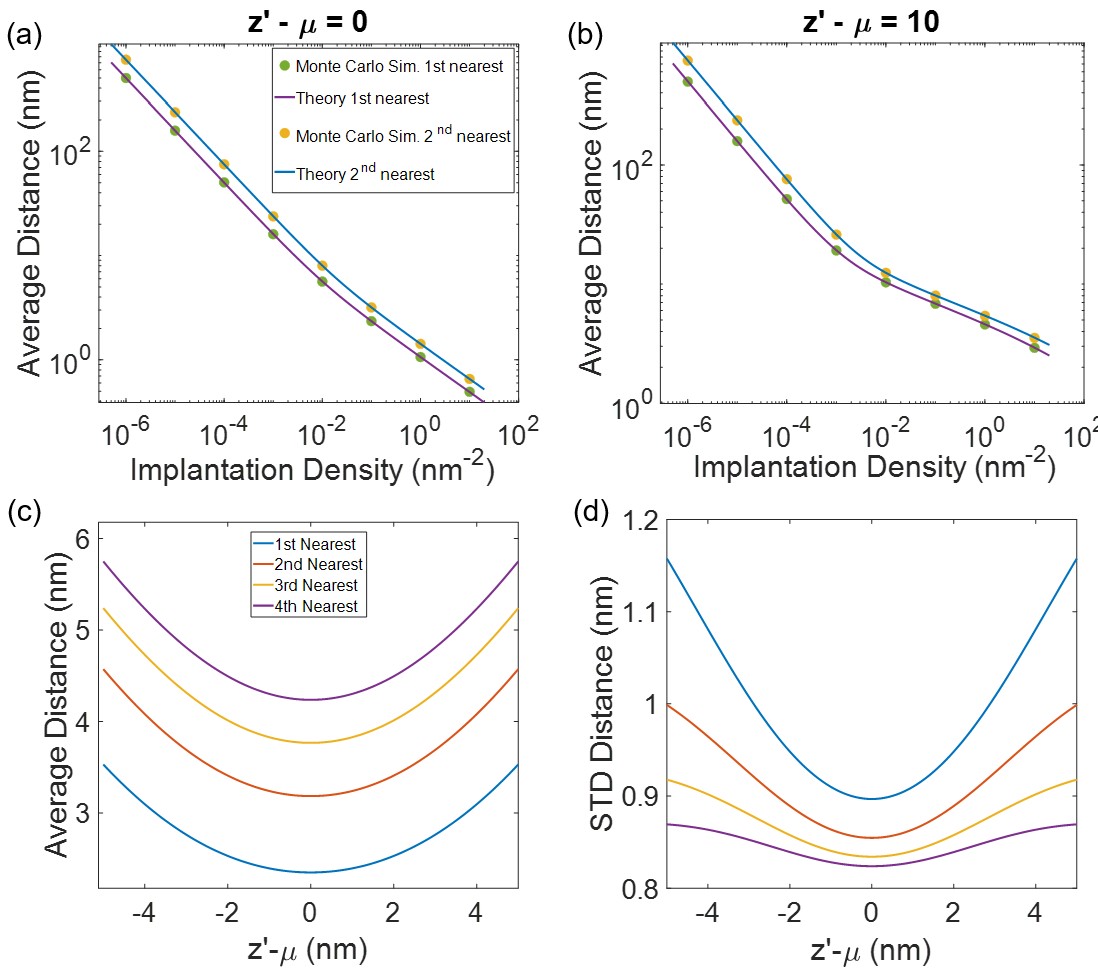}
\caption{Numerical and simulated $n_{\mathrm{th}}$ nearest-neighbor distance analysis. 
(a) NV center located at the mean depth $z' = \mu$. 
(b) NV center positioned 10~nm above the mean depth, i.e., $z' - \mu = 10$~nm. 
(c) Numerical results for distances up to the 4\textsuperscript{th} nearest neighbor. 
(d) Corresponding standard deviation for the data in (c).}
    \label{NV-P1 dis}
\end{figure}

Monte Carlo simulations were performed to verify the computed \( l_n(x', y', z') \). In the simulations, ions were assumed to be uniformly distributed along the \( x \)-\( y \) plane and Gaussian-distributed on the \( z \)-axis, with a standard deviation of 2.77 nm. Two cases were analyzed: (1) the NV center is exactly at the mean (\( |z' - \mu| = 0 \, \text{nm} \)), shown in Fig. \ref{NV-P1 dis}(a), and (2) the NV center is shifted 10 nm from the mean (\( |z' - \mu| = 10 \, \text{nm} \)), shown in Fig. \ref{NV-P1 dis}(b). As the ion density increased, both the nearest and second-nearest average distances decreased. When the ion density was low, the results for \( |z' - \mu| = 0 \, \text{nm} \) and \( |z' - \mu| = 10 \, \text{nm} \) were similar and aligned with the results of a 2D uniform distribution. This observation is sensible because, at low density, the uniform distribution along the \( x \)-\( y \) plane dominates, and the 2.77 nm variation along the \( z \)-axis can be ignored.

However, as the ion density increases, the typical distance along \( x \)-\( y \) becomes smaller. As a result, the variation along the \( z \)-axis starts to become significant, which causes deviations from the 2D uniform case. The \( |z' - \mu| = 0 \, \text{nm} \) case exhibited a smaller typical distance than the \( |z' - \mu| = 10 \, \text{nm} \) case because the ions were more dilute at locations farther from the mean. The values computed through our theory and simulations match very well, indicating our theory is accurate. 

To demonstrate the simplicity and comprehensiveness of our formalism, we plot up to the 4th nearest distance along with its standard deviation (Fig. \ref{NV-P1 dis}(c,d)). Additionally, our formalism allows for the computation of various types of probabilities, such as the probability of finding the $n^{th}$ nearest spin within a certain distance. For an NV center with random position, the probability density function for the NV is given by $G(x',y',z')$. Using this notation, the total average of the $n^{th}$ spin is given by 

\begin{equation}
    l_n = \iiint l_n(x',y',z')G(x',y',z')dx'dy'dz'.
\end{equation}

\noindent Notice that $G(x',y',z') = p(x',y',z')$ follows the same distribution as implanted nitrogen. The total variance of $l_n$ can be computed using the law of total variance 

\begin{equation}
\begin{split}
V_n=&\iiint V(x',y',z')G(x',y',z')dx'dy'dz'\\
&+\iiint l(x',y',z')^2G(x',y',z')dx'dy'dz\\
&-\bigg[\iiint l(x',y',z')G(x',y',z')dx'dy'dz\bigg]^2.
\end{split}
\end{equation}

\noindent The computed values are in the main text. This formalism can help with the implantation of ions and engineering design.

\section{Effect of dark spin decoherence}

The dark spins can undergo decoherence during the application of the $\pi$ pulse. As a result, the entire population of dark spins may not be flipped. Let $\alpha$ denote the fraction of dark spins successfully flipped by the $\pi$ pulse. These spins contribute to the DEER signal, while the remaining fraction $1-\alpha$, which are not flipped, contribute to the standard Hahn-echo background. 

The modified DEER signal, accounting for partial spin flipping, becomes

\begin{equation*}
    F_{\mathrm{DEER}}= e^{\alpha\sigma \iint f_{\mathrm{DEER}}-1dxdy+(1-\alpha)\sigma \iint f_{\mathrm{Echo}}-1dxdy}.
\end{equation*}

\noindent The Hahn-echo signal is unchanged

\begin{equation*}
    F_{\mathrm{Echo}}= e^{\sigma \iint f_{\mathrm{Echo}}-1dxdy}
\end{equation*}

\noindent and the ratio becomes:

\begin{equation}
   F= \frac{F_{\mathrm{DEER}}}{F_{\mathrm{Echo}}}= e^{\alpha \sigma \iint (f_{\mathrm{DEER}}-f_{\mathrm{Echo}})dxdy}.
\end{equation}

\noindent This expression shows that the sole effect of partial spin flipping (due to decoherence during the pulse) is to scale the effective estimated spin density by a factor of $\alpha$. In other words, dark spin decoherence reduces the apparent spin density, but does not otherwise alter the signal form. In our experiment, the microwave power and pulse durations are held constant throughout all measurements, ensuring the $\pi$ pulse duration and hence the value of $\alpha$ remains unchanged across all conditions.


\nocite{*}
\bibliography{TiO2_manuscript_bibliography}

\end{document}